\documentclass[twocolumn,english,reprint, preprintnumbers,nofootinbib,superscriptaddress]{revtex4}
\preprint{\bf YITP-SB-15-07}
\usepackage[hyperfootnotes=false,bookmarks=false]{hyperref}
\usepackage{graphicx}
\usepackage{amssymb}
\usepackage{amsmath}
\usepackage{multirow}
\usepackage{float}
\usepackage{color}
\newcommand{\as}{\alpha_s}

\begin{document}

\title{Hadronic production of $W$ and $Z$ bosons at large transverse momentum}

\author{Edmond L. Berger}
\affiliation{
High Energy Physics Division, Argonne National Laboratory, Argonne, Illinois 60439, USA}
\author{Jun Gao}
\affiliation{
High Energy Physics Division, Argonne National Laboratory, Argonne, Illinois 60439, USA}
\author{Zhong-Bo Kang}
\affiliation{
Theoretical Division, Los Alamos National Laboratory, Los Alamos, NM 87545, USA}
\author{Jian-Wei Qiu}
\affiliation{
Physics Department, Brookhaven National Laboratory, Upton, NY 11973-5000, USA}
\affiliation{
C.N.~Yang Institute for Theoretical Physics and Department of Physics and Astronomy, 
Stony Brook University, Stony Brook, NY 11794-3840, USA}
\author{Hao Zhang}
\affiliation{Department of Physics, University of California, Santa Barbara, CA 93106, USA}

\begin{abstract}

We introduce a modified factorization formalism in quantum chromodynamics for 
hadronic production of $W$ and $Z$ bosons at large transverse momentum 
$p_T$. When $p_T$ is much larger than the invariant mass $Q$ of the vector boson, 
this new factorization formalism systematically resums the large fragmentation logarithms, 
$\as^m\ln^m(p_T^2/Q^2)$, to all orders in the strong coupling $\as$.  
Using our modified factorization formalism, we present 
the next-to-leading order (NLO) predictions for $W$ and $Z$ boson production 
at high $p_T$ at the CERN Large Hadron Collider and at a future 100 TeV proton-proton collider.
Our NLO results are about $5\%$ larger in normalization, and they show improved convergence and 
moderate reduction 
of the scale variation compared to the NLO predictions derived in a conventional fixed-order 
perturbative expansion.\\


\end{abstract}

\maketitle

\section{Introduction}
\label{sec:int}

Successful operation of the CERN Large Hadron Collider (LHC) and associated particle detectors have 
led to the discovery of the Higgs boson, the final piece of the standard model 
(SM)~\cite{Chatrchyan:2012ufa,Aad:2012tfa} of particle physics, along with the exploration of strong 
interaction short-distance phenomena in high 
energy processes at much greater values of the production transverse momentum.  Future experimental 
investigations, at higher collision energy and with greater luminosity, promise refined understanding 
of the nature of electroweak symmetry breaking and possible evidence of new physics beyond the 
SM.  Some of these searches will focus on deviations from SM expectations or on anomalies in 
high-energy tails of various kinematic distributions.  Precise SM predictions for these observables and 
distributions are important assets for discovery of new physics.

The production distributions of massive electroweak (EW) gauge bosons, $W$'s and $Z$'s, are among 
observables that can be sensitive to physics beyond the SM, either because new states in extensions of the 
SM may decay into 
$W$ and $Z$ bosons, or because the lepton distributions from SM $W$ and $Z$ decay are important
backgrounds for high-energy lepton signatures in new physics models.  The leptons could mimic the 
signature of boosted objects from the decay of a new heavy resonance.  Moreover, $W$ and $Z$ 
boson production serve as tests of perturbative quantum chromodynamics (QCD) calculations, 
and data on their distributions are important in the determination of the parton distribution functions 
(PDFs)~\cite{Berger:1988tu}. 
Among measurements of $W$ and $Z$ production, precise cross sections at large transverse momentum 
hold particular interest.  The SM $Z$ boson production at large $p_T$ can be used for jet energy-scale 
calibration.   Existing studies show the possibility of using $W$ and $Z$ boson production at large $p_T$ 
to further constrain the gluon PDFs~\cite{Kim:1990kt,Malik:2013kba}.  Transverse momentum distributions 
of $W$ and $Z$ bosons have been measured at the LHC~\cite{Aad:2014xaa,Aad:2014haa} but with 
relatively large uncertainties.  

To exploit the full potential of the data, ever more precise SM predictions are needed, 
requiring better understanding of the size of high-order perturbative corrections.  
The $p_T$ spectra of $W$ and $Z$ boson production in hadronic collisions have been calculated 
perturbatively in the SM.  The leading order (LO) cross section at a finite $p_T$ is of ${\mathcal O}(\alpha_{em}\as)$. 
The next-to-leading order (NLO) QCD corrections were calculated decades 
ago~\cite{Gonsalves:1989ar,Baer:1991qf,Arnold:1990yk}, and found to be sizable for the
energy of the LHC.  The EW corrections were studied extensively in recent 
years~\cite{Maina:2004rb,Kuhn:2004em,Kuhn:2005az,Hollik:2007sq,Kuhn:2007qc,Kuhn:2007cv,Becher:2013zua,Kallweit:2014xda}.   
The $W$ and $Z$ production at small $p_T\ll Q$ ($\sim$ mass of $W$ or $Z$) has received a lot of attention 
in connection with the resummation of the effects Sudakov double logarithms in 
QCD~\cite{Collins:1984kg,Davies:1984hs,Balazs:1997xd,Ellis:1997ii,Qiu:2000ga,Qiu:2000hf,Berger:2002ut,Landry:2002ix,Mantry:2010mk,Becher:2010tm,Bozzi:2010xn,Mantry:2010bi,Becher:2011xn,Catani:2013tia,Wang:2013qua}, 
also important for precise measurements of the $W$ boson mass in hadronic collisions.  
Improved predictions are awaited of $W$ and $Z$ production at large $p_T$ beyond NLO 
corrections~\cite{Kidonakis:1999ur,Kidonakis:2003xm,Gonsalves:2005ng,Becher:2011fc,Kidonakis:2012sy,Becher:2012xr,Becher:2013vva,Kidonakis:2014zva}.   
Work is in progress on the full next-to-next-to-leading order (NNLO) QCD corrections to 
the $p_T$ spectrum of $W$ and $Z$ production, similar to the case of Higgs
boson production~\cite{Boughezal:2013uia,Chen:2014gva}
\footnote{After our paper was submitted, the NNLO QCD corrections to
$W+$ jet production appeared~\cite{Boughezal:2015dva}}.  

\begin{figure}[h!]
  \begin{center}
  \includegraphics[width=0.22\textwidth]{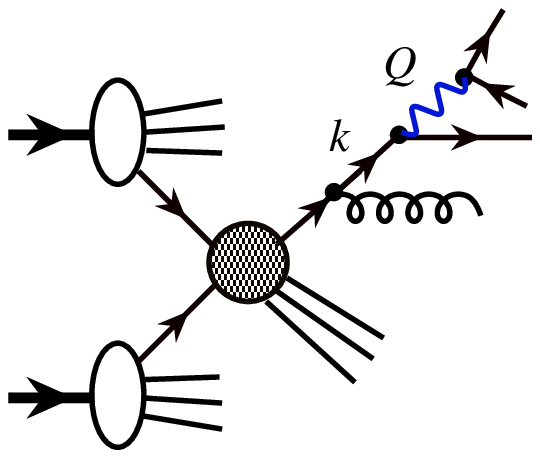}\hspace{0.1in}
  \includegraphics[width=0.22\textwidth]{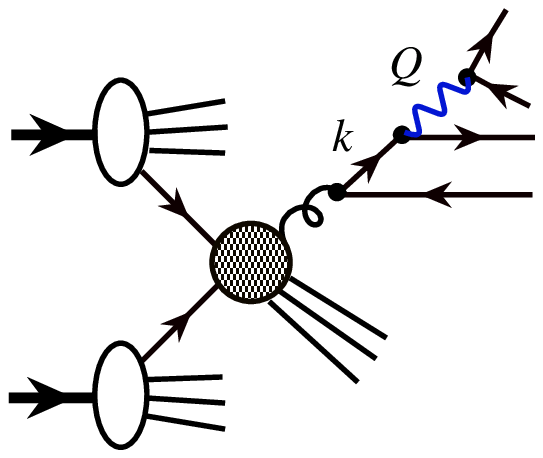}
  \end{center}
  \vspace{-1ex}
  \caption{\label{fig:fflogs}
Leading order QCD diagrams that lead to the fragmentation logarithms in $W$ and $Z$ production. 
The diagram on the left (right) provides the fragmentation logarithm from quark (gluon) splitting. } 
\end{figure}

In this paper, we explore another approach to theoretical improvement by identifying potentially large QCD logarithms 
from high order perturbative calculations, and resumming these logarithms to all orders.  
QCD corrections to the short-distance partonic scattering cross sections of $W$ and $Z$ 
production at large $p_T$ could receive one power of a large logarithm, $\ln(p_T^2/Q^2)$, 
for every additional power of $\as$.   Such large logarithms come from partonic subprocesses 
in which the high $p_T$ heavy boson is radiated from a more energetic quark 
(or a parton in general), as is illustrated in Fig.~\ref{fig:fflogs}.  
For the production of EW gauge bosons with mass $Q\sim 100$~GeV, the fragmentation logarithm, 
$\ln(p_T^2/Q^2) \sim 4.6$, when the boson's transverse momentum $p_T\sim 1$~TeV, and 
could be a potential source of large high order corrections. 

The leading contribution from the
fragmentation logarithm of the diagrams in Fig.~\ref{fig:fflogs} has the approximate form,   
\begin{eqnarray}
\hat{\sigma}^{\rm F} / \hat{\sigma}_f^{\rm LO}
&\sim &
\frac{\alpha_s(\mu)}{2\pi}\, P_{f\to q}(z)\int_{k^2_{\rm min}}^{k^2_{\rm max}} \frac{dk^2}{k^2}\
\nonumber \\
&\sim & 
\frac{\alpha_s(\mu)}{2\pi}\, \ln\left(\frac{p_T^2}{Q^2}\right)  ,
\label{eq:flogs}
\end{eqnarray}
where $\hat{\sigma}_f^{\rm LO}$, with parton flavor $f=q,g$, represents 
the lower order contribution to the production cross section 
without the quark to quark-gluon (or gluon to quark-antiquark) splitting in Fig.~\ref{fig:fflogs}, 
and $P_{f\to q}(z)$ is the LO splitting function 
for a parton of flavor $f$ to fragment into a quark, 
e.g., $P_{g\to q}(z) = (1/2)[z^2 + (1-z)^2]$ with the color factor 1/2.  
To obtain the second line in Eq.~(\ref{eq:flogs}), 
we use the fact that the fragmentation contribution is dominated by the large $z$ region
and the size of the LO splitting functions to a quark at large $z$ is of ${\cal O}(1)$.
Since the factorization scale could be chosen from ${\cal O}(Q)$ to ${\cal O}(p_T)$,  
we could estimate the size of the high order corrections from the LO fragmentation
logarithms as  
\begin{equation}
\frac{\alpha_s(\mu)}{2\pi}\ln\left(\frac{p_T^2}{Q^2}\right) 
\lesssim 
\frac{\alpha_s(Q)}{2\pi}\ln\left(\frac{p_T^2}{Q^2}\right) 
\lesssim 10\%\, ,
\label{eq:fsize}
\end{equation}
for $p_T\sim 1$~TeV.  
That is, the higher order corrections from the fragmentation logarithms to the production
of heavy EW gauge boson of mass $Q\sim 100$~GeV, 
$[(\alpha_s(\mu)/2\pi)\ln(p_T^2/Q^2)]^m$ with $m>1$, 
should be under control perturbatively. It was pointed out in~\cite{Rubin:2010xp}
that the double logarithms from radiation of soft $Z$ bosons can
induce large higher-order corrections in the large jet $p_T$ region of
$Z+$ jets production. This situation is different from the case
studied here where we are looking at inclusive $Z$ boson production
at large $p_T$. The double logarithms are highly suppressed in
our case.    

In this paper, we use a modified QCD factorization formalism for calculating the hadronic cross 
sections of $W$ and $Z$ production at large transverse momentum $p_T$.  This formalism is similar 
to the one that we introduced for the low mass Drell-Yan cross section at large $p_T$ 
in previous work~\cite{Berger:1998ev,Berger:2001wr,Kang:2008wv}. 
In order to resum the $\as^m\ln^m(p_T^2/Q^2)$-type large logarithms, we re-organize the conventional 
fixed-order perturbative 
expansion of $W$ and $Z$ cross sections at large $p_T$ into factored ``direct'' and 
``fragmentation'' contributions, as demonstrated in Sec.~\ref{sec:frm}: 
\begin{equation}
\frac{d\sigma_{AB\rightarrow V(Q) X}}{dp_T^2\,dy}\equiv
\frac{d\sigma^{\rm Dir}_{AB\rightarrow V(Q) X}}{dp_T^2\,dy}+
\frac{d\sigma^{\rm Frag}_{AB\rightarrow V(Q) X}}{dp_T^2\,dy}. \nonumber
\end{equation}
All powers of the logarithmic $\as\ln(p_T^2/Q^2)$ high order corrections 
are completely resummed into $W$ and $Z$ fragmentation functions, and 
included into the ``fragmentation'' contributions, while the short-distance 
partonic hard parts for both ``direct'' and ``fragmentation'' contributions are
free of the large logarithms, and can be systematically calculated order-by-order
in powers of $\as$.  Using our modified factorization scheme, we calculate
the next-to-leading order (NLO) predictions for $W$ and $Z$ boson production 
at high $p_T$ at LHC energies, as well as at a future 100 TeV proton-proton collider.
The perturbatively calculated $W$ and $Z$ cross sections at large $p_T$ show improved 
convergence, as well as moderate reduction of the scale variations compared to 
the conventional NLO perturbative expansion.

The situation in $W$ and $Z$ production is very different from the fragmentation contributions 
to light hadron production, where the logarithms could run into the non-perturbative region, and 
also numerically different from the role of fragmentation contributions to the production
of low mass Drell-Yan pairs or heavy quarkonia at high $p_T$ \cite{Berger:2001wr,Ma:2014svb}.  
The key difference is the mass of the heavy EW gauge bosons.
In order to produce a heavy EW gauge bosons of mass $Q$, as shown in Fig.~\ref{fig:fflogs}, 
the invariant mass of the fragmenting parton of momentum $k$ should be sufficiently large, 
as $\sqrt{k^2} \gtrsim Q$.  Consequently, the radiation from the fragmenting parton is similar 
to the radiation from a heavy quark of mass $\sqrt{k^2} \gtrsim Q$, and 
is strongly suppressed for the phase space within the angle, $\sqrt{k^2}/k_0$, 
of the fragmenting parton, the so-called ``dead-cone'' effect~\cite{Azimov:1982ef}.  In addition to 
the much smaller 
phase space for the radiation, the large virtuality of the fragmenting parton also sets 
the renormalization scale $\mu$ for the strong coupling constant $\alpha_s(\mu) \lesssim
\alpha_s(M_Z) \sim 0.118$ with $Z$ mass, $M_Z$.  With the current limit of collision energies, it is the 
combination of small $\alpha_s$ and the restricted phase space for the radiation that 
controls the size of the corrections from high order fragmentation logarithms.
In this paper, we verify this conclusion by performing explicit all order resummation of the fragmentation logarithms.

The rest of our paper is organized as follows. In Sec.~\ref{sec:frm}, we 
introduce our modified factorization scheme for EW gauge boson production at large 
$p_T$ and compare it with the conventional fixed-order perturbative expansion scheme. 
In Sec.~\ref{sec:num}, applying our modified factorization scheme, 
we present our predictions for the $W$ and $Z$ fragmentation functions, 
as well as our calculations of the LO and NLO cross sections of $W$ and $Z$ 
production at large $p_T$ at LHC energies and at a $\sqrt{s}=100$ TeV
future proton-proton collider.  We also discuss the improvement of 
our modified factorization formalism over the conventional fixed-order
perturbative expansion.  Our summary and conclusions are presented in 
Sec.~\ref{sec:con}.

\section{QCD Factorization of vector boson production at large $p_T$}
\label{sec:frm}

The cross section for producing an on-shell EW gauge boson of momentum $p^\mu$ 
can be factored systematically in QCD perturbation theory
as~\cite{Collins:1989gx}
\begin{eqnarray}
\frac{d\sigma_{AB\rightarrow V(Q) X}}{dp_T^2\,dy}
&&=\sum_{a,b}\int dx_1 f_a^A(x_1,\mu) 
           \int dx_2 f_b^B(x_2,\mu)\, \nonumber \\
&&\times
\frac{d\hat{\sigma}^{\rm Pert}
                    _{ab\rightarrow V(Q) X}}{dp_T^2\,dy}
 (x_1,x_2,Q,p_T,y;\mu) \, ,
\label{Vph-fac}
\end{eqnarray}
under the usual assumption that the physically measured quantities $Q$ and $p_T$ 
are both much larger than $\Lambda_{\rm QCD}$. 
In Eq.~(\ref{Vph-fac}), the variables $Q$, $p_T$, and $y$
are the mass, transverse momentum, and rapidity of the
vector boson, respectively; and the symbol $X$ stands 
for an inclusive sum over final states that recoil against 
the observed vector boson. 
The sum $\sum_{a,b}$ runs over all parton flavors, and
$f_a^A$ and $f_b^B$ are the corresponding PDFs,
with the partons' momentum fractions $x_1$ and $x_2$, respectively; 
$\mu$ represents the renormalization and factorization scale, which  
is of the order of the energy exchange of the hard collision: 
$\mu \sim \sqrt{Q^2 + p_T^2}$.  
The function $d\hat{\sigma}^{\rm Pert}_{ab\rightarrow V(Q) X}/dp_T^2 dy$ 
in Eq.~(\ref{Vph-fac}) is the short-distance hard part of the partonic scattering cross section.  
It can be calculated in perturbation theory 
in powers of the QCD coupling $\alpha_s(\mu)$.   
The first two terms of the perturbative expansion,  
$d\hat{\sigma}^{\rm C-LO}$ and $d\hat{\sigma}^{\rm C-NLO}$, including 
perturbative contributions up to ${\mathcal O}(\alpha_{em}\alpha_s)$ and
${\mathcal O}(\alpha_{em}\alpha_s^2)$ respectively, have been available 
in the literature for sometime \cite{Gonsalves:1989ar,Baer:1991qf,Arnold:1990yk}.  
The superscript ``${\rm C}$'' indicates the {\em conventional} fixed-order 
perturbative expansion.  

Beyond the leading order in $\alpha_s$, when $p_T \gg Q$, the
perturbative functions $d\hat{\sigma}^{\rm Pert}_{ab\rightarrow V(Q) X}/dp_T^2 dy$ 
in Eq.~(\ref{Vph-fac}) can receive large high order corrections 
in powers of $\alpha_s\ln(p_T^2/Q^2)$ arising from the radiation 
of partons along the direction of the observed vector boson.  
Such large logarithmic corrections can be
systematically resummed into the parton to vector-boson 
fragmentation functions, $D_{f\to V}$, as demonstrated for
the case of the virtual photon~\cite{Berger:2001wr}.
The perturbative series for 
$d\hat{\sigma}^{\rm Pert}
              _{ab\rightarrow V(Q) X}/dp_T^2 dy$
can therefore be re-organized into two terms as in~\cite{Berger:2001wr},
\begin{eqnarray}
\frac{d\hat{\sigma}^{\rm Pert}_{ab\rightarrow V(Q) X}}
     {dp_T^2\,dy}(x_1,x_2,Q,p_T,y;\mu)
= &&\nonumber \\
&&\hspace{-1.7in}\frac{d\hat{\sigma}^{\rm Dir}_{ab\rightarrow V(Q) X}}
     {dp_T^2\,dy}(x_1,x_2,Q,p_T,y;\mu,\mu_D)
\nonumber \\  
&&\hspace{-1.7in}+
\frac{d\hat{\sigma}^{\rm Frag}_{ab\rightarrow V(Q) X}}
     {dp_T^2\,dy}(x_1,x_2,Q,p_T,y;\mu,\mu_D)\, , 
\label{fac-dir-frag}
\end{eqnarray}
where the superscripts ``Dir'' and ``Frag'' represent the
``direct'' and the ``fragmentation'' contribution, respectively.
The latter includes the perturbative 
fragmentation logarithms, and it can be further factored 
as \cite{Qiu:2001nr}, 
\begin{eqnarray}
\frac{d\hat{\sigma}^{\rm Frag}_{ab\rightarrow V(Q) X}}
     {dp_T^2 dy}
&=& \sum_{c} \int \frac{dz}{z^2}\, 
  \left[
  \frac{d\hat{\sigma}_{ab\rightarrow c X}}
       {dp_{c_T}^2\,dy}\left(x_1,x_2,p_c;\mu_D\right)
  \right] \nonumber \\ 
 && \times D_{c\rightarrow V}(z,\mu_D^2;Q^2) ,
\label{DY-F}
\end{eqnarray}
where $p_c^\mu=\hat{p}^\mu/z$, 
with $\hat{p}^\mu$ 
defined to be $p^\mu$ 
at $Q^2=0$, corresponding to the approximation $Q^2\ll p_T^2$ made
for
$d\hat{\sigma}_{ab\rightarrow c X}/dp_{c_T}^2\,dy$ in Eq.~(\ref{DY-F}),
which is a short-distance hard part for partons of flavors 
$a$ and $b$ to produce a parton of flavor $c$ and 
momentum $p_c$. The fragmentation scale $\mu_D\sim p_T$
is introduced to separate the direct and fragmentation contributions 
in Eq.~(\ref{fac-dir-frag}), and its dependence would be cancelled
if both contributions include all order corrections~\cite{Berger:2001wr}.

The fragmentation contribution in Eq.~(\ref{DY-F}) 
shares the typical two-stage generic pattern of the
fragmentation production of a single particle at
large transverse momentum $p_T$ 
(much larger than its mass): the production of an on-shell
parton of flavor $c$ at the distance scale $1/p_T$, convoluted with
a fragmentation function that includes the leading logarithmic
contributions from the ``running'' of the distance scale 
from $1/\mu_D\sim 1/p_T$ to $1/\mu_{D0}\sim 1/Q$. 
When $Q\gg \Lambda_{\rm QCD}$, which is the case for the $W$ and $Z$ boson,
the parton to vector-boson fragmentation function $D_{f\to V}$ is perturbative, and so is
the whole resummed fragmentation contribution, 
$\hat{\sigma}^{\rm Frag}$ in Eq.~(\ref{DY-F})~\cite{Berger:2001wr}.

The direct contribution in Eq.~(\ref{fac-dir-frag}) is 
perturbatively calculable in a power series of $\alpha_s$
\cite{Berger:2001wr}, and it is defined as, 
\begin{equation}
\frac{d\hat{\sigma}^{\rm Dir}_{ab\rightarrow V(Q) X}}{dp_T^2\,dy}\equiv
\frac{d\hat{\sigma}^{\rm Pert}_{ab\rightarrow V(Q) X}}{dp_T^2\,dy}-
\frac{d\hat{\sigma}^{\rm F-Asym}_{ab\rightarrow V(Q) X}}{dp_T^2\,dy},
\label{DY-D}
\end{equation}
where $d\hat{\sigma}^{{\rm Pert}}$ is
the conventional fixed-order perturbative QCD calculation.  Moreover, 
$d\hat{\sigma}^{{\rm F-Asym}}$, 
with the superscript ``Asym'' referring to an ``asymptotic'' contribution,
is simply the perturbative 
expansion of the resumed fragmentation contribution, 
$d\hat{\sigma}^{\rm Frag}$ in Eq.~(\ref{DY-F}), to the same order 
in powers of $\alpha_s$ as $d\hat{\sigma}^{{\rm Pert}}$ in Eq.~(\ref{DY-D}).
It is effectively a subtraction term to systematically remove from 
$d\hat{\sigma}^{{\rm Pert}}$ all fragmentation logarithms which have 
been resummed into the fragmentation contribution, 
$d\hat{\sigma}^{\rm Frag}$ in Eq.~(\ref{DY-F}).  The subtraction  
avoid double counting order-by-order in powers of $\alpha_s$.  
Consequently, the direct contribution is free of 
the large fragmentation logarithms, while it still 
keeps all non-logarithmic terms from 
the physics between the scales $p_T$ and $Q$. 
Note that $d\hat{\sigma}^{\rm Dir-LO}=d\hat{\sigma}^{\rm C-LO}$
since the fragmentation contributions start at ${\mathcal O}(\alpha_{em}\as^2)$.

By substituting Eqs.~(\ref{fac-dir-frag}) and (\ref{DY-F}) into Eq.~(\ref{Vph-fac}), we obtain
our modified factorization formalism for heavy EW vector boson production 
in hadronic collisions,
\begin{equation}
\frac{d\sigma_{AB\rightarrow V(Q) X}}{dp_T^2\,dy}\equiv
\frac{d\sigma^{\rm Dir}_{AB\rightarrow V(Q) X}}{dp_T^2\,dy}+
\frac{d\sigma^{\rm Frag}_{AB\rightarrow V(Q) X}}{dp_T^2\,dy},
\label{eq:modified-fac}
\end{equation}
where $d\sigma^{\rm Dir}$ is factored in the same way as $d\sigma$
in Eq.~(\ref{Vph-fac}) with $d\hat{\sigma}^{\rm Pert}$ replaced by 
$d\hat{\sigma}^{\rm Dir}$ defined in Eq.~(\ref{DY-D}), and 
$d\sigma^{\rm Frag}$ is 
\begin{eqnarray}
\frac{d\sigma^{\rm Frag}_{AB\rightarrow V(Q) X}}{dp_T^2\,dy}
&=&\sum_{a,b,c}\int dx_1 f_a^A(x_1,\mu) 
           \int dx_2 f_b^B(x_2,\mu)\, \nonumber \\
&&
\times \int \frac{dz}{z^2}\, 
  \left[
  \frac{d\hat{\sigma}^{\rm Frag}_{ab\rightarrow c X}}
       {dp_{c_T}^2\,dy}\left(x_1,x_2,p_c;\mu_D\right)
  \right] \nonumber \\ 
&&
\times \,
D_{c\rightarrow V}(z,\mu_D^2;Q^2),
\label{DR-F-fac}
\end{eqnarray}
where the fragmentation functions $D_{c\rightarrow V}(z,\mu_D^2;Q^2)$
resum all fragmentation logarithms, and
$d\hat{\sigma}^{\rm Frag}_{ab\rightarrow c X}$ are 
partonic hart parts for producing an on-shell parton of flavor ``$c$'' and 
momentum $p_c^\mu$, which are independent of the specific vector
boson produced.  Actually, they are the same as the perturbative coefficient 
functions for producing a light hadron, such as pion, and are
available for both the LO and NLO in powers of $\alpha_s$ in the literature 
\cite{Aversa:1988vb}.

The separation between the direct and the fragmentation contribution 
in Eq.~(\ref{eq:modified-fac}) depends on the definition of
the parton to vector-boson fragmentation functions.  
Different definitions of the fragmentation functions correspond to a
different scheme to split the conventional fixed-order perturbative expansion
into the ``direct" plus ``fragmentation" contributions.
A scheme choice for the fragmentation functions should also fix
the asymptotic contribution and, correspondingly, the direct contribution.
The sum of these two terms in our modified factorization formalism 
in Eq.~(\ref{eq:modified-fac}) should not be very sensitive to the scheme choice.

The key difference between our modified factorization formula in Eq.~(\ref{eq:modified-fac})
and the conventional factorization formula in Eq.~(\ref{Vph-fac})
resides in the way the large logarithmic contributions
from final-state parton splitting are handled. Instead of
one perturbative series in powers of $\alpha_s$ in the
conventional approach, we have two perturbative expansions
in our modified factorization formula: one for the direct
and one for the fragmentation contribution. All coefficient
functions in the new perturbative expansions are free of
large logarithms. The large perturbative logarithms in the conventional
expansion in powers of $\alpha_s$ are systematically 
resummed into the fragmentation functions.
As a result, the perturbative expansion of the partonic hard parts 
in our modified factorization approach has better convergence
properties than the conventional fixed-order expansion. 
In addition, because of the reorganization and resummation of the logarithms,
both the LO and NLO cross sections in our new approach 
include a tower of logarithmically enhanced high order 
corrections from the conventional fixed-order perturbative 
expansion.  For example, 
for the LO cross section, with the lowest order partonic hard parts retained, 
the difference between our modified approach in Eq.~(\ref{eq:modified-fac}) 
and the conventional fixed-order approach in Eq.~(\ref{Vph-fac}) is 
\begin{eqnarray}
&&
\sum_{a,b}\int {\hskip -0.04in} dx_1 f_a^A {\hskip -0.05in}
                 \int {\hskip -0.04in} dx_2 f_b^B {\hskip -0.05in}
\left\{
\left[ \frac{d\hat{\sigma}^{\rm D-LO}_{ab\rightarrow V(Q) X}}
               {dp_{T}^2\,dy}
     - \frac{d\hat{\sigma}^{\rm C-LO}_{ab\rightarrow V(Q) X}}
              {dp_{T}^2\,dy}
       \right] \right. 
       \nonumber\\
&&\hspace{1.2in} + 
\left.
      \sum_{c} \int {\hskip -0.04in} \frac{dz}{z^2}\, 
      \frac{d\hat{\sigma}^{\rm F-LO}_{ab\rightarrow c X}}
       {dp_{c_T}^2\,dy}\, D_{c\rightarrow V(Q)}
        \right\}
      \nonumber \\
&&
=\sum_{a,b,c}\int {\hskip -0.04in} dx_1 f_a^A {\hskip -0.05in}
                \int {\hskip -0.04in} dx_2 f_b^B {\hskip -0.05in}
              \int {\hskip -0.04in} \frac{dz}{z^2}\,
              \frac{d\hat{\sigma}^{\rm F-LO}_{ab\rightarrow c X}}
              {dp_{c_T}^2\,dy}\, D_{c\rightarrow V(Q)}    ,        
\label{eq:diff-lo}
\end{eqnarray}
where the LO relation, $d\hat{\sigma}^{\rm Dir-LO}=d\hat{\sigma}^{\rm C-LO}$,
is used.  Similarly, at NLO, the difference is 
\begin{eqnarray}
&&
\sum_{a,b}\int {\hskip -0.04in} dx_1 f_a^A {\hskip -0.05in}
                 \int {\hskip -0.04in} dx_2 f_b^B {\hskip -0.05in}
\left\{
\left[ \frac{d\hat{\sigma}^{\rm D-NLO}_{ab\rightarrow V(Q) X}}
               {dp_{T}^2\,dy}
     - \frac{d\hat{\sigma}^{\rm C-NLO}_{ab\rightarrow V(Q) X}}
              {dp_{T}^2\,dy}
       \right] \right. 
       \nonumber\\
&&\hspace{1.2in} 
+ \left.
      \sum_{c} \int {\hskip -0.04in} \frac{dz}{z^2}\, 
      \frac{d\hat{\sigma}^{\rm F-NLO}_{ab\rightarrow c X}}
       {dp_{c_T}^2\,dy}\, D_{c\rightarrow V(Q)}
        \right\}
\nonumber \\
&&
=\sum_{a,b,c}\int {\hskip -0.03in} dx_1 f_a^A {\hskip -0.04in}
                 \int {\hskip -0.03in} dx_2 f_b^B {\hskip -0.04in}
                 \int {\hskip -0.03in} \frac{dz}{z^2}\,
\nonumber\\
&&  \hspace{0.4in}
\times \left\{
\frac{d\hat{\sigma}^{\rm F-LO}_{ab\rightarrow c X}}
                        {dp_{c_T}^2\,dy}
\left[D_{c\rightarrow V(Q)}  - D^{(0)}_{c\rightarrow V(Q)} \right]
\right.
\nonumber\\  
&&  \hspace{0.5in}
\left.
+
\left[  \frac{d\hat{\sigma}^{\rm F-NLO}_{ab\rightarrow c X}}
                        {dp_{c_T}^2\,dy}
       - \frac{d\hat{\sigma}^{\rm F-LO}_{ab\rightarrow c X}}
                        {dp_{c_T}^2\,dy}
                        \right] 
                        D_{c\rightarrow V(Q)}
\right\} .
\label{eq:diff-nlo} 
\end{eqnarray}
In deriving Eq.~(\ref{eq:diff-nlo}), we use the definition in Eq.~(\ref{DY-D}) and 
the following relation, 
\begin{equation}
\frac{d\hat{\sigma}^{\rm F-Asym-NLO}_{ab\rightarrow V(Q) X}}
              {dp_{T}^2\,dy}
= \sum_c \int {\hskip -0.03in} \frac{dz}{z^2}\,
   \frac{d\hat{\sigma}^{\rm F-LO}_{ab\rightarrow c X}}
                        {dp_{c_T}^2\,dy} 
                        D^{(0)}_{c\rightarrow V(Q)}
\label{eq:fasym-nlo}
\end{equation}
where $D^{(0)}_{c\to V(Q)}$ is the fragmentation function for a single parton of flavor $c$ 
to fragment into a vector boson of invariant mass $Q$, expanded perturbatively
to ${\cal O}(\alpha_{em}\alpha_s^0)$.

In comparison with a conventional fixed-order calculation at the same orders,  
Eqs.~(\ref{eq:diff-lo}) and (\ref{eq:diff-nlo}) show that 
our modified factorization formalism in Eq.~(\ref{eq:modified-fac}) 
includes the all orders resummation of final-state fragmentation 
logarithms, plus additional short-distance high order contributions.  
The first term on the right-hand side of Eq.~(\ref{eq:diff-nlo}), 
comes from the resummation of leading logarithms to all orders, and 
the second term is from the ${\mathcal O}(\alpha_s^3)$ corrections 
to the partonic hard parts.  To the order in which we work, the full correction 
is the sum of the two.  In our presentation of numerical results 
in Sec.~\ref{sec:num},  the sum of the two terms in Eq.~(\ref{eq:diff-nlo})
is referred to as the type-B correction to the NLO term 
of the conventional fixed-order perturbative calculation, 
while the first term alone is denoted the type-A correction. 

The resummation of the fragmentation logarithms into the single
parton to a vector-boson fragmentation functions is achieved by solving
the inhomogeneous evolution equations \cite{Qiu:2001nr}, 
\begin{eqnarray}
\mu_D^2 \frac{d}{d\mu_D^2} D_{c\rightarrow V}(z,\mu_D^2;Q^2)
&{\hskip -0.1in} = {\hskip -0.1in}& 
\left(\frac{\alpha_{\mathrm{em}}}{2\pi}\right)
 \gamma_{c\rightarrow V}(z,\mu_D^2,\alpha_s;Q^2)
\nonumber \\
&&\hspace{-1.45in}+
\left(\frac{\alpha_{s}}{2\pi}\right)
\sum_d \int_z^1 \frac{dz'}{z'}
P_{c\rightarrow d}(\frac{z}{z'},\alpha_s)\, 
 D_{d\rightarrow V}(z',\mu_D^2;Q^2),
 \nonumber \\
\label{RG-unpol}
\end{eqnarray}
where $c,d=q,\bar{q},g$.  The ambiguity in defining the 
fragmentation function is connected to the renormalization of its perturbative 
ultraviolet (UV) divergence, and
the choice of the fragmentation scale, $\mu_D$.  
In Eq.~(\ref{RG-unpol}), the evolution kernels 
$P_{c\rightarrow d}$ are infrared (IR) safe, evaluated at a single hard scale,
$\mu_D$, and calculated perturbatively as a power series in $\alpha_s$.   
These kernels are insensitive to the specific vector meson produced, and 
are the same as the evolution kernels for pion or other light hadron 
fragmentation functions.  They are the same as the splitting functions of the
Dokshitzer-Gribov-Lipatov-Altarelli-Parisi (DGLAP) evolution
equations~\cite{Collins:1988wj} at the leading order. 

The inhomogeneous term in the evolution equations can
also be calculated perturbatively, and in general it has
power correction terms of the form $Q^2/\mu_D^2$, owing to the 
mass of the vector boson, and therefore, it depends on the nature of the vector
boson produced.  In the invariant mass cutoff
scheme~\cite{Qiu:2001nr,Braaten:2001sz}, the lowest order 
quark-to-virtual photon evolution kernel $\gamma_{q\to\gamma^*}^{(0)}$ 
was derived in \cite{Qiu:2001nr}.  With a simple generalization, we
compute the inhomogeneous evolution kernel for a quark to fragment into 
an EW gauge boson of invariant mass $Q$, $\gamma_{q\to V}^{(0)}$, 
and we obtain,
\begin{eqnarray}
\gamma_{q\rightarrow V}^{(0)}(z,k^2;Q^2) 
&=&
\frac{(|g_L^{Vq}|^2+|g_R^{Vq}|^2)}{2} \left[\frac{1+(1-z)^2}{z}\right. 
\nonumber\\
&&{\hskip 0.2in}
\left.  -z\left(\frac{Q^2}{zk^2}\right) \right]
      \theta(k^2-\frac{Q^2}{z})\, ,
\label{Gq0-unpol}
\end{eqnarray}
where $k^2$ is the invariant mass of the parent quark and 
is identified as $\mu_D^2$, and the $\theta$-function is 
a consequence of the mass threshold.  Here, $g_{L,R}^{Vq}$ are the
EW couplings between quarks and the EW gauge bosons, 
with $\{g_L^{Wq},\,g_R^{Wq}\}
=\{1/(\sqrt 2s_w),\,0\}$ for the $W$ boson, and
$\{g_L^{Zq},\,g_R^{Zq}\}=\{(1/2-2s_w^2/3)/(s_wc_w),\,-2s_w/(3c_w)\}$ and
$\{(-1/2+s_w^2/3)/(s_wc_w),\,+s_w/(3c_w)\}$ for the $Z$ boson to couple to
the up and down-type quarks respectively.  We us 
$s_w$ and $c_w$ to represent the sine and cosine of the weak-mixing angle.
The gluon to vector-boson evolution kernel vanishes 
at the lowest order 
\begin{equation}
\gamma_{g\rightarrow V}^{(0)}(z,k^2;Q^2) 
= 0\, ,
\label{Gg0-unpol}
\end{equation}
because the gluon does not interact directly with the EW gauge bosons. 

The choice of factorization scheme is not unique.  We use the invariant mass
cutoff scheme~\cite{Qiu:2001nr,Braaten:2001sz}. 
Some choices, such as the modified minimum subtraction ($\overline{\rm MS}$) 
scheme,  may not respect the mass threshold when $Q^2\neq 0$ and 
lead to negative fragmentation functions 
\cite{Qiu:2001nr,Braaten:2001sz}. 
QCD corrections to the lowest order parton to vector-boson  
splitting function $\gamma_{c\rightarrow V}^{(0)}$ can be evaluated 
in principle order-by-order in $\alpha_s$.

If $Q\gg \Lambda_{\rm QCD}$,
the parton to vector-boson fragmentation functions are completely 
perturbative, as is the case for our study here.  
The lowest order parton to vector-boson fragmentation functions are
\begin{eqnarray}
&&{\hskip -0.1in}
D_{q\to V}^{(0)}(z,\mu_D^2;Q^2)
=\frac{(|g_L^{Vq}|^2+|g_R^{Vq}|^2)}{2}
\left(\frac{\alpha_{em}}{2\pi}\right)
\nonumber\\
&&
\times\left[
\frac{1+(1-z)^2}{z}\ln\left(\frac{z\mu_D^2}{Q^2}\right)
-z\left(1-\frac{Q^2}{z\mu_D^2}\right)\right],
\label{eq:Dq0}\\
&&{\hskip -0.1in}
D_{g\to V}^{(0)}(z,\mu_D^2;Q^2) = 0\, .
\label{eq:Dg0}
\end{eqnarray}
We must specify a boundary condition in order to solve the evolution equations 
in Eq.~(\ref{RG-unpol}).   A natural boundary 
condition following the mass threshold constraint is~\cite{Qiu:2001nr,Berger:2001wr}
\begin{equation}
D_{c\rightarrow V}(z,\mu_D^2\le Q^2/z;Q^2) = 0\, ,
\label{pert_input}
\end{equation}
for any flavor $c$, if we choose the invariant mass cutoff scheme for
the fragmentation functions.

To conclude this section, we emphasize that our modified
factorization formalism in Eq.~(\ref{fac-dir-frag}) effectively reorganizes
the {\it single} perturbative expansion of conventional
QCD factorization, in Eq.~(\ref{Vph-fac}), into {\it two} perturbative
expansions in Eq.~(\ref{eq:modified-fac}), 
plus the perturbatively calculated evolution kernels for the fragmentation functions.
The main advantage of this reorganization is that the
new perturbative expansions are evaluated at a single hard
scale and are free of large logarithms.  As shown in 
Sec.~\ref{sec:num}, the ratios of the NLO over the LO contributions
in the new perturbative expansions are smaller than the
ratios evaluated in the conventional approach.

\section{Numerical results}
\label{sec:num}

We present our numerical results in this section. We choose the $G_F$
parametrization scheme~\cite{Denner:1990ns} for the EW
couplings with $M_W=80.385\,{\rm GeV}$, $M_Z=91.1876\,{\rm GeV}$, $M_{t}=173\,{\rm GeV}$,
and $G_F=1.166379\times 10^{-5}\,{\rm GeV}^{-2}$~\cite{Beringer:1900zz}.
We assume a diagonal Cabibbo-Kobayashi-Maskawa matrix of the SM in the calculation
for simplicity. We use META1.0 PDF~\cite{Gao:2013bia} of the proton, which is a NNLO PDF set
that combines information from the CT10~\cite{Gao:2013xoa}, MSTW2008~\cite{Martin:2009iq}, 
and NNPDF2.3~\cite{Ball:2012cx} PDF sets,
and three-loop running of the QCD
coupling constants with $N_f=5$ active quark flavors and $\alpha_s(M_Z)=0.118$.

\subsection{Fragmentation functions}

At LO only quarks can fragment into EW vector bosons, for which the fragmentation
functions are proportional to the square of the EW couplings.  As seen in Eq.~(\ref{eq:Dq0}), 
for example, the up-type quark can fragment into a $W^+$ boson at LO, 
but not into a $W^-$ boson. The fragmentation functions are
the same for quark and anti-quark to virtual-photon or $Z$ boson. Similarly
fragmentation functions are equal for up-type quarks to $W^+$ and down-type
quarks to $W^-$. The gluon can fragment into
$W$ or $Z$ bosons only at higher orders through intermediate quarks, thus the fragmentation
functions are suppressed.

\begin{figure}[h!]
  \begin{center}
  \includegraphics[width=0.45\textwidth]{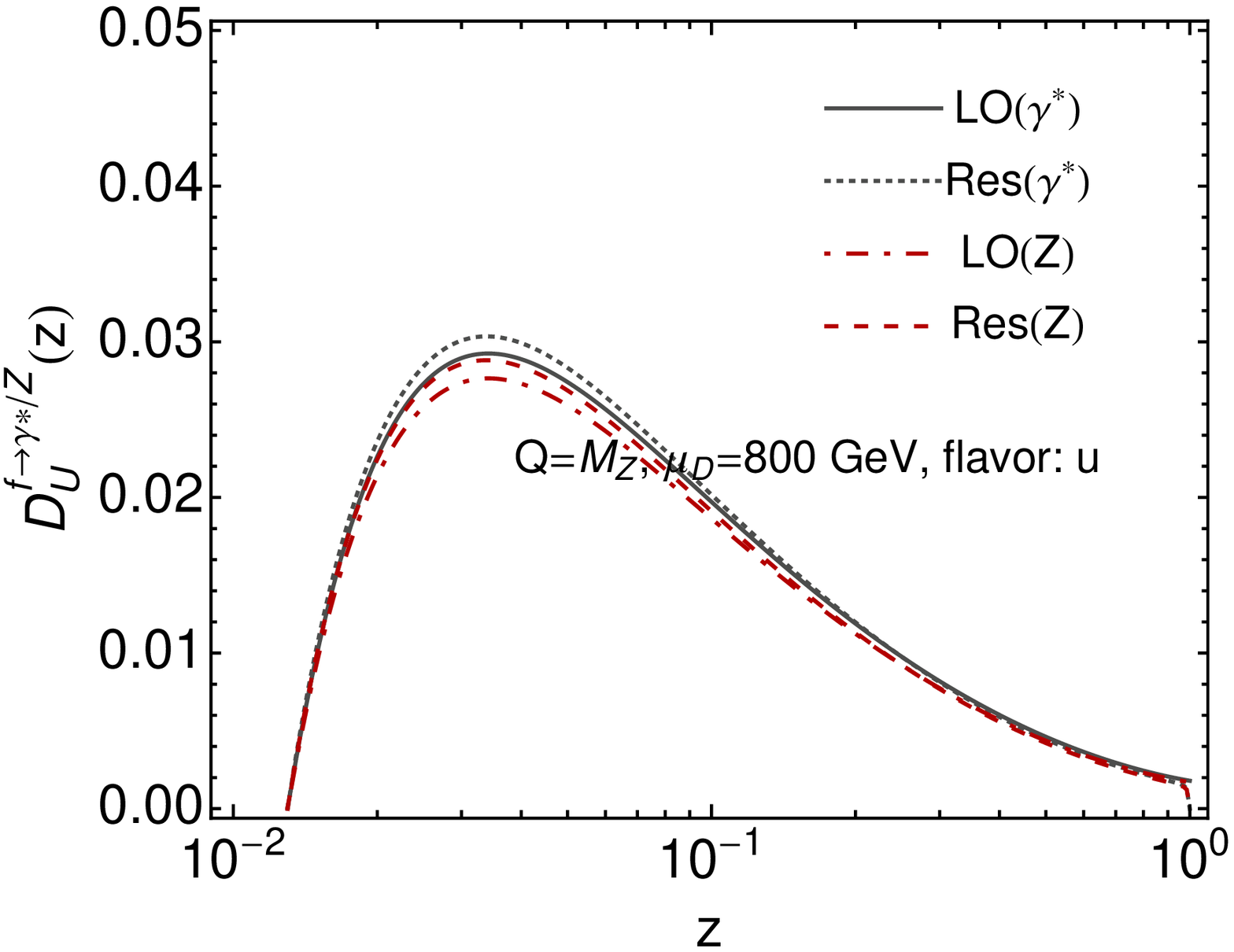}\vspace{0.1in}
  \includegraphics[width=0.45\textwidth]{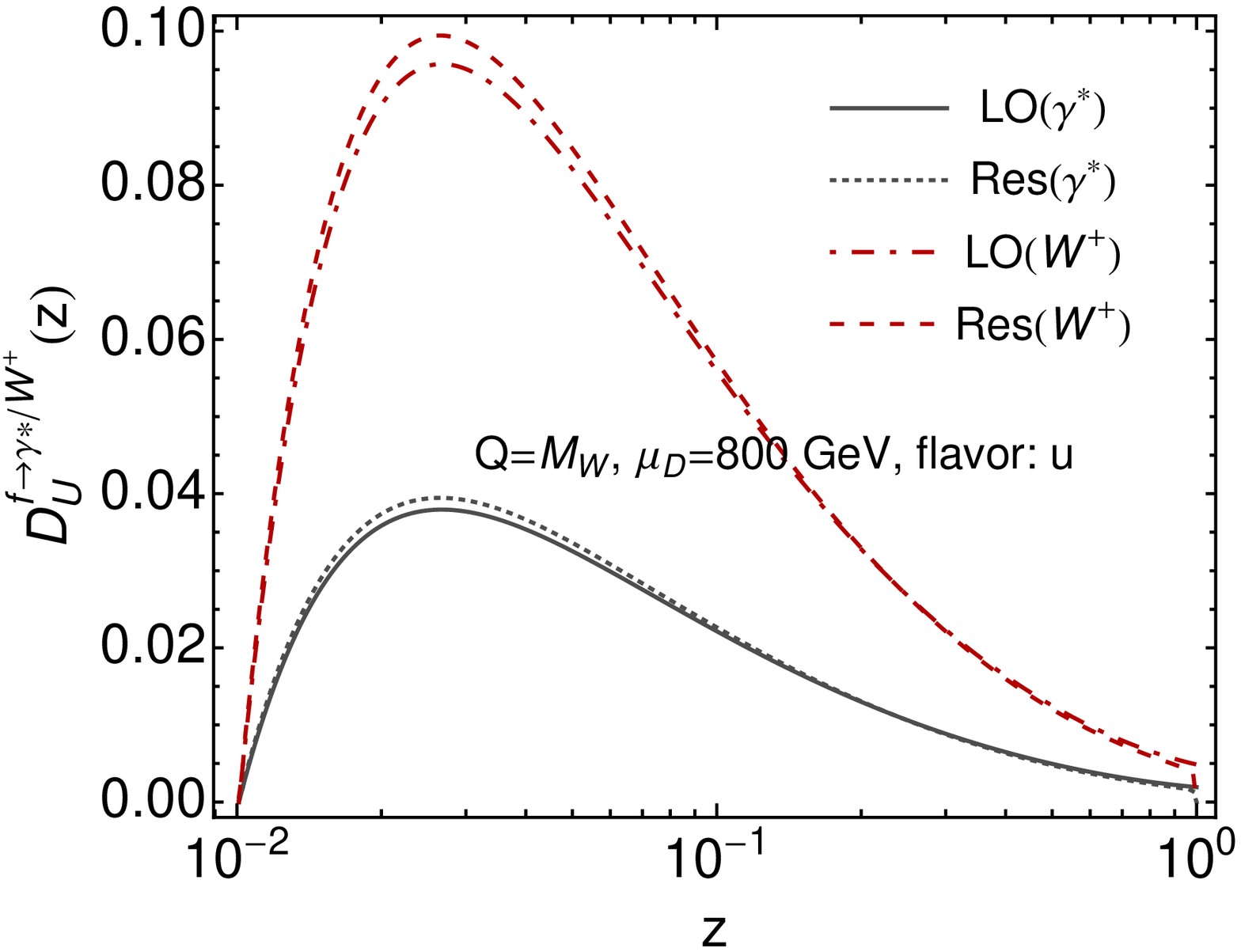}
  \end{center}
  \vspace{-1ex}
  \caption{\label{fig:frag1}
 The $z$ dependence of the LO and resummed fragmentation functions of a $u$ quark into 
 a $Z$ and a $W^{+}$ boson for a fragmentation scale $\mu_D=800\,{\rm GeV}$.
 For reference, we also plot the fragmentation function of a virtual photon with
 the same mass.} 
\end{figure}

\begin{figure}[h!]
  \begin{center}
  \includegraphics[width=0.45\textwidth]{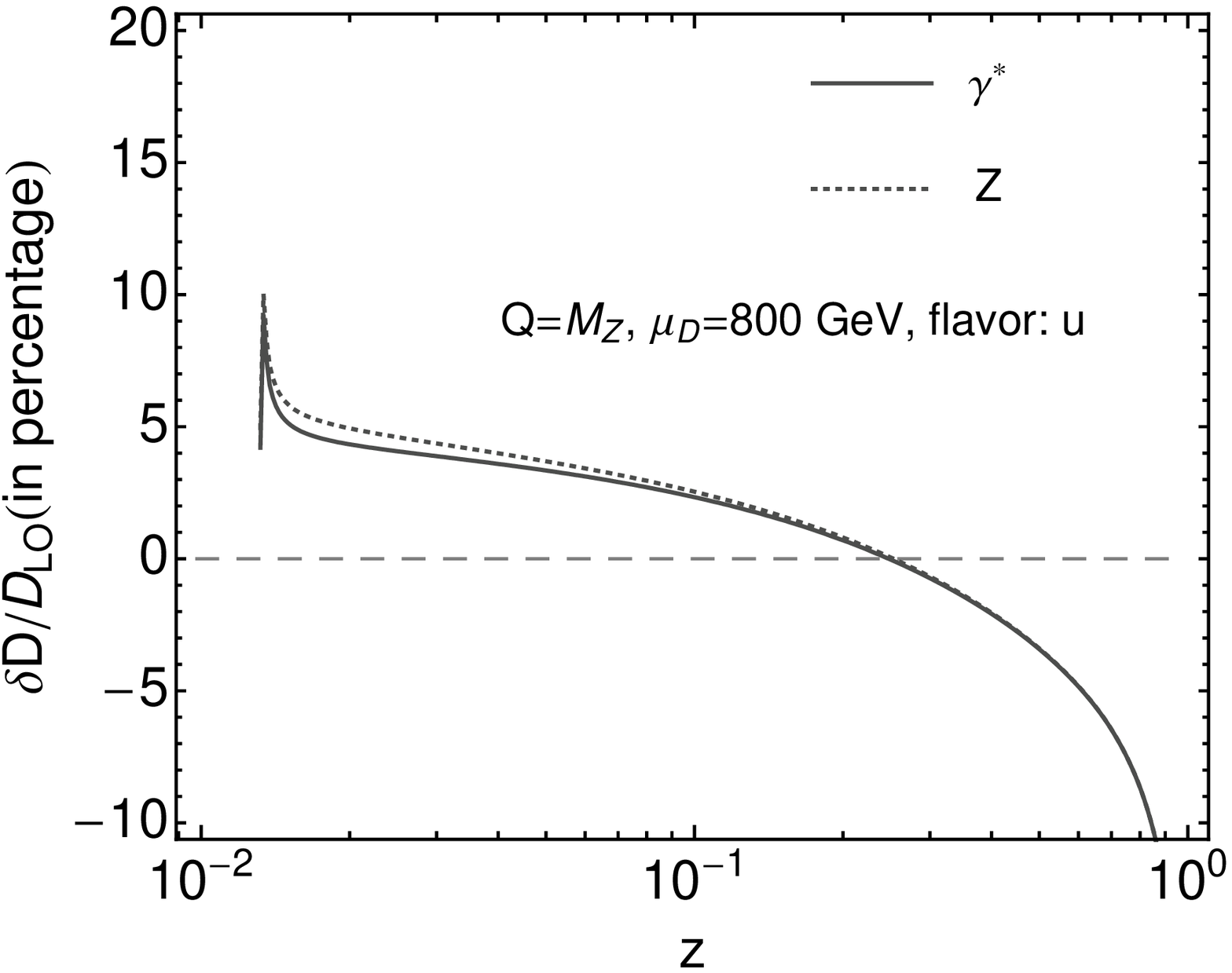}\vspace{0.1in}
  \includegraphics[width=0.45\textwidth]{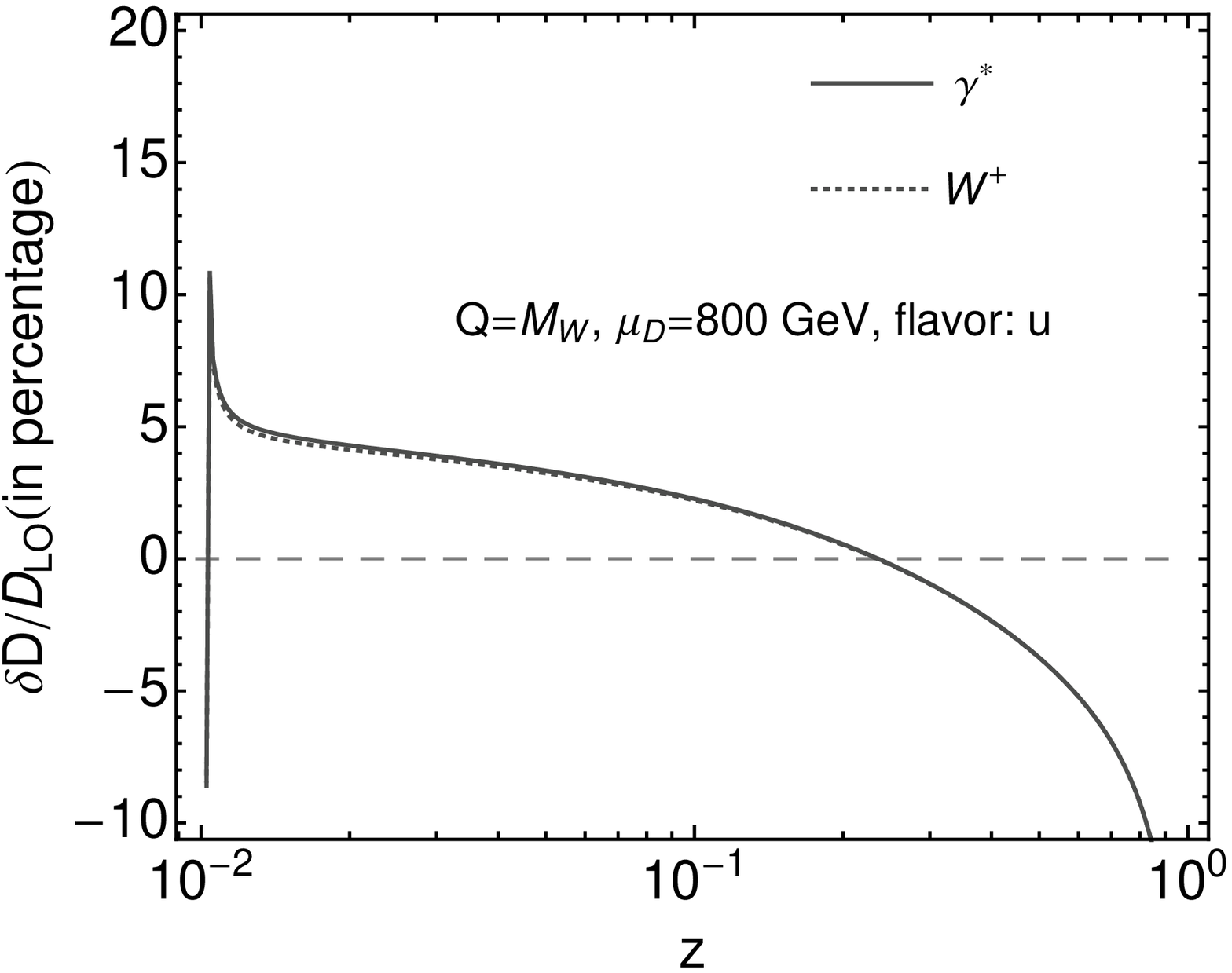}
  \end{center}
  \vspace{-1ex}
  \caption{\label{fig:frag2}
 Percentage change of the resummed to the LO fragmentation functions for a $u$ quark
 to (a) $Z$ and (b) $W^{+}$ boson for a fragmentation scale $\mu_D=800\,{\rm GeV}$.} 
\end{figure}
In Fig.~\ref{fig:frag1}, we show the momentum fraction $z$ dependence of the LO and resummed parton 
to vector-boson fragmentation functions for the $Z$ and $W^{+}$ bosons.   In
each figure we also plot the parton to virtual-photon fragmentation functions with the same
mass.  We choose a typical fragmentation scale of 800 $\rm GeV$
and show results for the $u$ quark as an example. The fragmentation
functions have a peak in $z$, a consequence of the vector boson's mass threshold.   
The height of the peak is proportional to the square of the corresponding EW couplings.

\begin{figure}[h!]
  \begin{center}
  \includegraphics[width=0.45\textwidth]{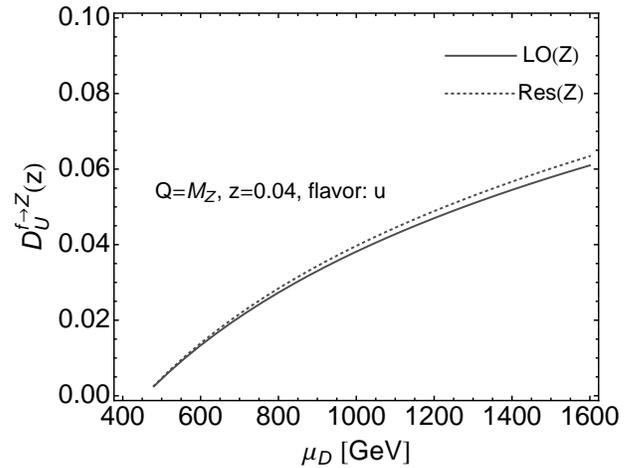}
  \end{center}
  \vspace{-1ex}
  \caption{\label{fig:frag4}
  Evolution of the LO and resummed fragmentation functions for a $u$ quark
  into a $Z$ boson with momentum fraction $z=0.04$.}
\end{figure}

In Fig.~\ref{fig:frag2}, we show the resummed corrections to the LO fragmentation functions
in percentage.  As expected,
the resummed contributions increase the fragmentation function at small and
moderate $z$ values while they reduce it for large $z$ values.
The observed differences of the fractional corrections for different
vector bosons are caused by the difference of their EW couplings through
the mixing of singlet and non-singlet evolutions. We further show
the evolution of the fragmentation function for a $u$ quark into
a $Z$ boson with momentum fraction $z=0.04$ in Fig.~\ref{fig:frag4}.
The scale dependence of the LO fragmentation functions are dominated
by the logarithmic term in Eq.~(\ref{eq:Dq0}). The resummed corrections
here are always positive. 

\subsection{$W$ and $Z$ boson
production at large $p_T$ in the conventional expansion scheme}

\begin{figure}[h!]
  \begin{center}
  \includegraphics[width=0.45\textwidth]{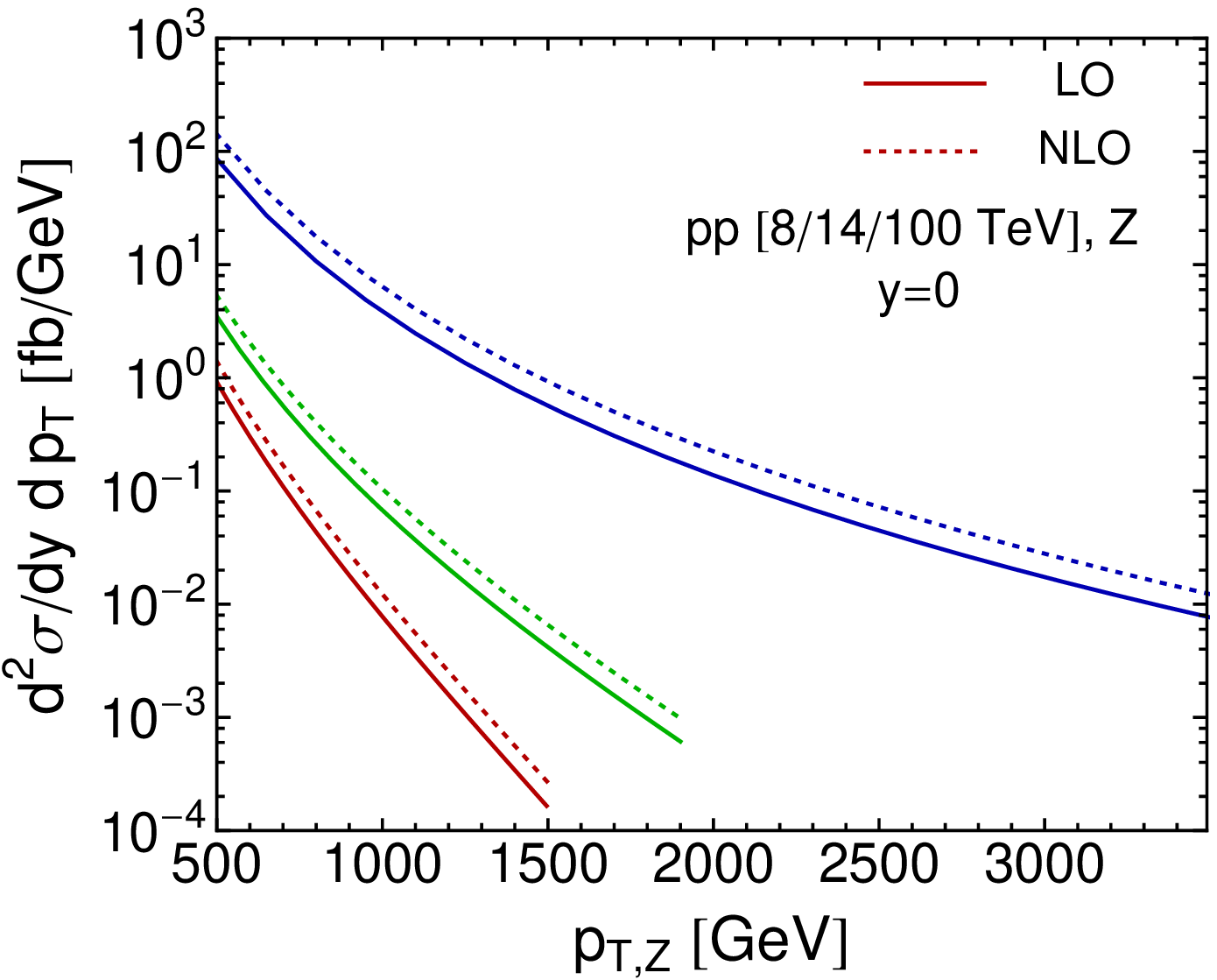}\vspace{0.1in}
  \includegraphics[width=0.45\textwidth]{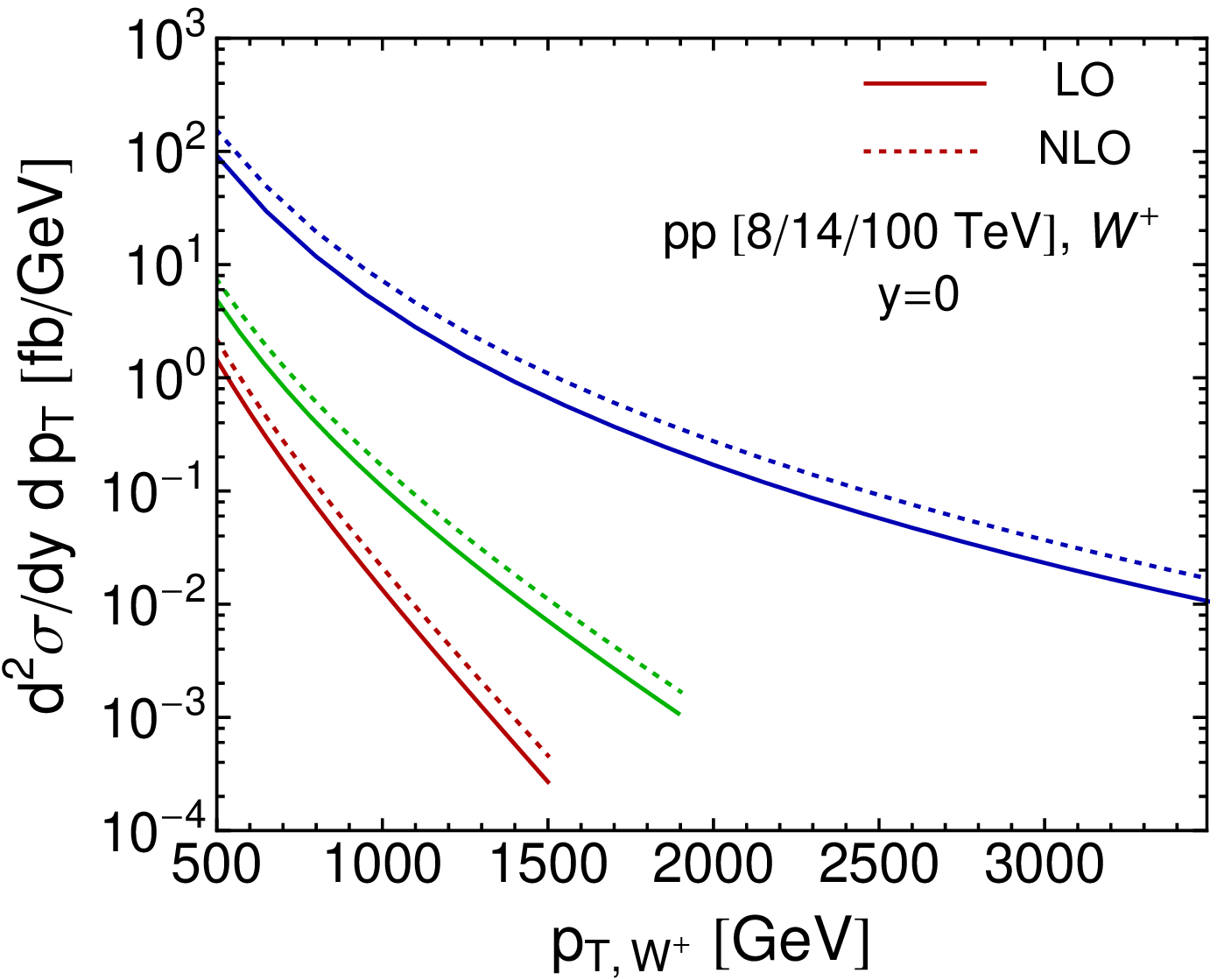}
  \end{center}
  \vspace{-1ex}
  \caption{\label{fig:xsec1}
 The LO and NLO predictions for the double-differential cross sections of
 $Z$ and $W^+$ boson production at large transverse momentum, $p_T$,
 in the conventional fixed-order perturbative expansion, 
 at $\sqrt{s}= 8$ and 14 TeV at the LHC, and 100 TeV at a future proton-proton collider.   
 The range of $p_T$ is determined by the available phase space of the collisions. }
\end{figure}

In this subsection we present the cross sections of $W$ and $Z$ boson production
at large $p_T$ at LO and NLO accuracy in the conventional fixed-order perturbative 
expansion, defined in Eq.~(\ref{Vph-fac}), based on the analytical results
in~\cite{Gonsalves:1989ar,Baer:1991qf,Arnold:1990yk}.
We plot in Fig.~\ref{fig:xsec1} the $p_T$ dependence of the double-differential cross sections 
of the vector bosons at a central rapidity, $y=0$.
We show the results at $\sqrt{s}=8$ and 14 TeV at the LHC, 
and at $\sqrt{s}=100$~TeV at a future proton-proton collider, with
increasing range of $p_T$ as allowed by the higher collision energy.
We focus on the large $p_T$ region ($>500\,{\rm GeV}$) where the
perturbative logarithms, $\propto \ln(p_T/Q)$, are large.
We set the factorization and renormalization scales, as well as the fragmentation scale,
equal to the transverse mass $m_T=\sqrt{Q^2+p_T^2}$, unless specified otherwise.
The scale dependence of our numerical results is discussed later.
As shown in Fig.~\ref{fig:xsec1} the cross sections decrease quickly with increasing
$p_T$ at the LHC, caused by the fast drop of the partonic flux at large partonic  
momentum fraction $x$.  With a larger phase space available 
at the 100 TeV collider, the $p_T$ spectrum decreases more slowly. 
The NLO QCD corrections can be as large as 50\% of the LO cross sections 
for the LHC energies, and reach 60\% for the 100 TeV collider.

\begin{figure}[h!]
  \begin{center}
  \includegraphics[width=0.45\textwidth]{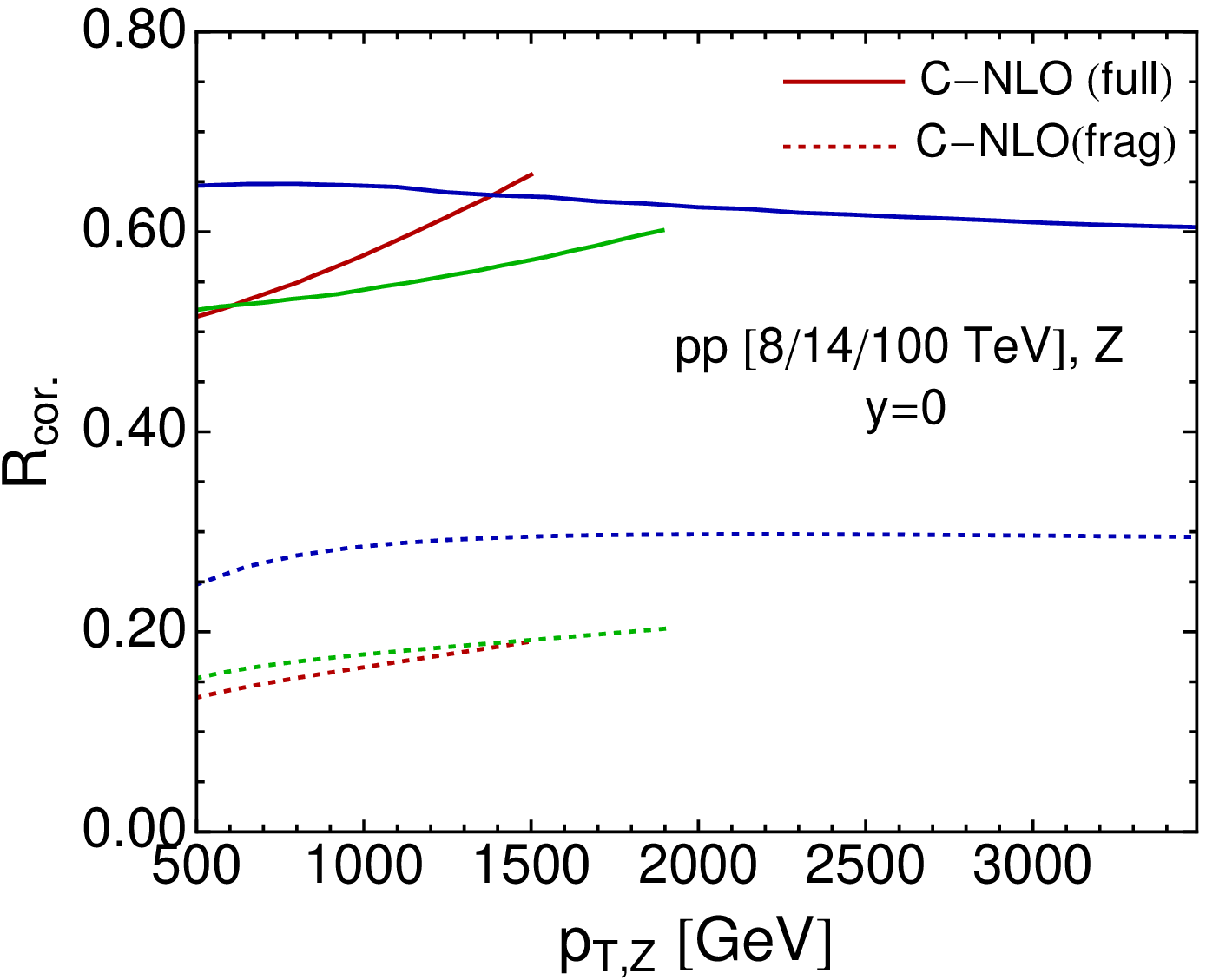}\vspace{0.1in}
  \includegraphics[width=0.45\textwidth]{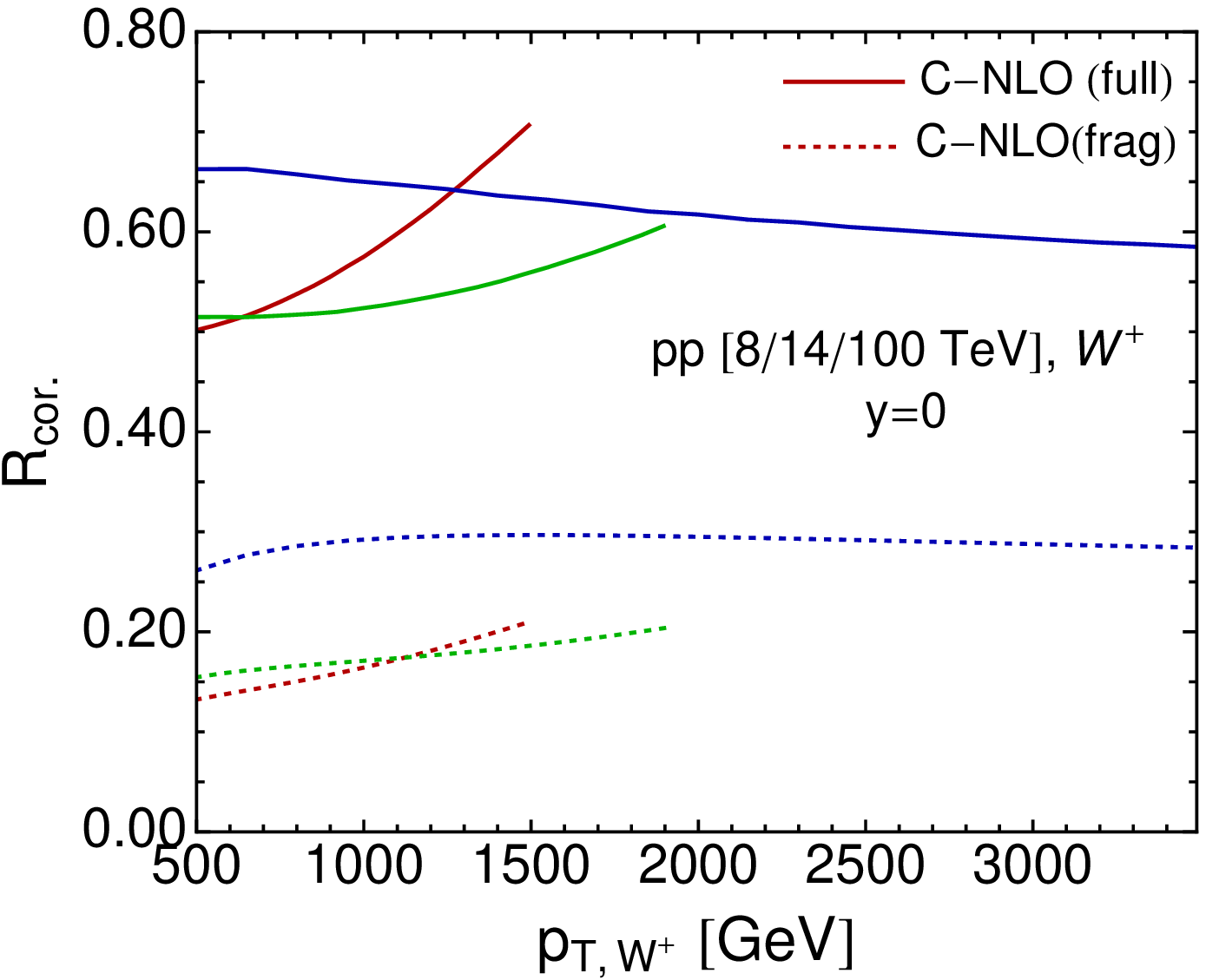}
  \end{center}
  \vspace{-1ex}
  \caption{\label{fig:rat1}
 The full NLO corrections in the conventional fixed-order expansion scheme,
 and the corresponding contributions proportional to the fragmentation logarithms, both 
 normalized to the conventional LO cross sections, are plotted as a function of  
 $p_T$ for $Z$ and $W^+$ boson production in proton-proton collisions at $\sqrt{s}=8$, 14 and 100 TeV. }
\end{figure}

To quantify the size of contributions from the large fragmentation logarithms 
we compare the full NLO corrections with the part of the corrections 
proportional to the fragmentation logarithms, 
equal to the fragmentation contribution in Eq.~(\ref{DR-F-fac}) expanded to the same
order in $\as$.   In Fig.~\ref{fig:rat1}, we plot the normalized full NLO corrections,
${\rm R}_{\rm cor.} = (\sigma^{\rm NLO} - \sigma^{\rm LO})/\sigma^{\rm LO}$ 
(solid lines), along with the normalized contributions proportional 
to the fragmentation logarithms (dashed lines).
The full NLO corrections are large
and increase quickly for vector-boson production at the LHC as 
$p_T$ approaches its maximum.  The behavior differs slightly for different vector bosons
owing to the different masses and different contributing PDFs.
The fragmentation pieces contribute about one third of the full
NLO corrections for the $p_T$ values considered. 
With the larger phase space available for radiation, 
the NLO fragmentation contributions are more significant at the 100 TeV collider. 
Their magnitude is as large as 30\% of the LO cross sections 
or about one half of the full NLO corrections. 
The fragmentation contributions are almost insensitive to 
the value of $p_T$ for the 100 TeV collider, and they stay pretty much
constant when $p_T\gtrsim 1$ TeV for both $W$ and $Z$ bosons.
This result comes about because the logarithmic enhancements
from $\ln(p_T/Q)$ are also modulated by the $z$ dependence 
of the hard partonic cross sections in the convolution with the
fragmentation functions.

\begin{figure}[h!]
  \begin{center}
  \includegraphics[width=0.45\textwidth]{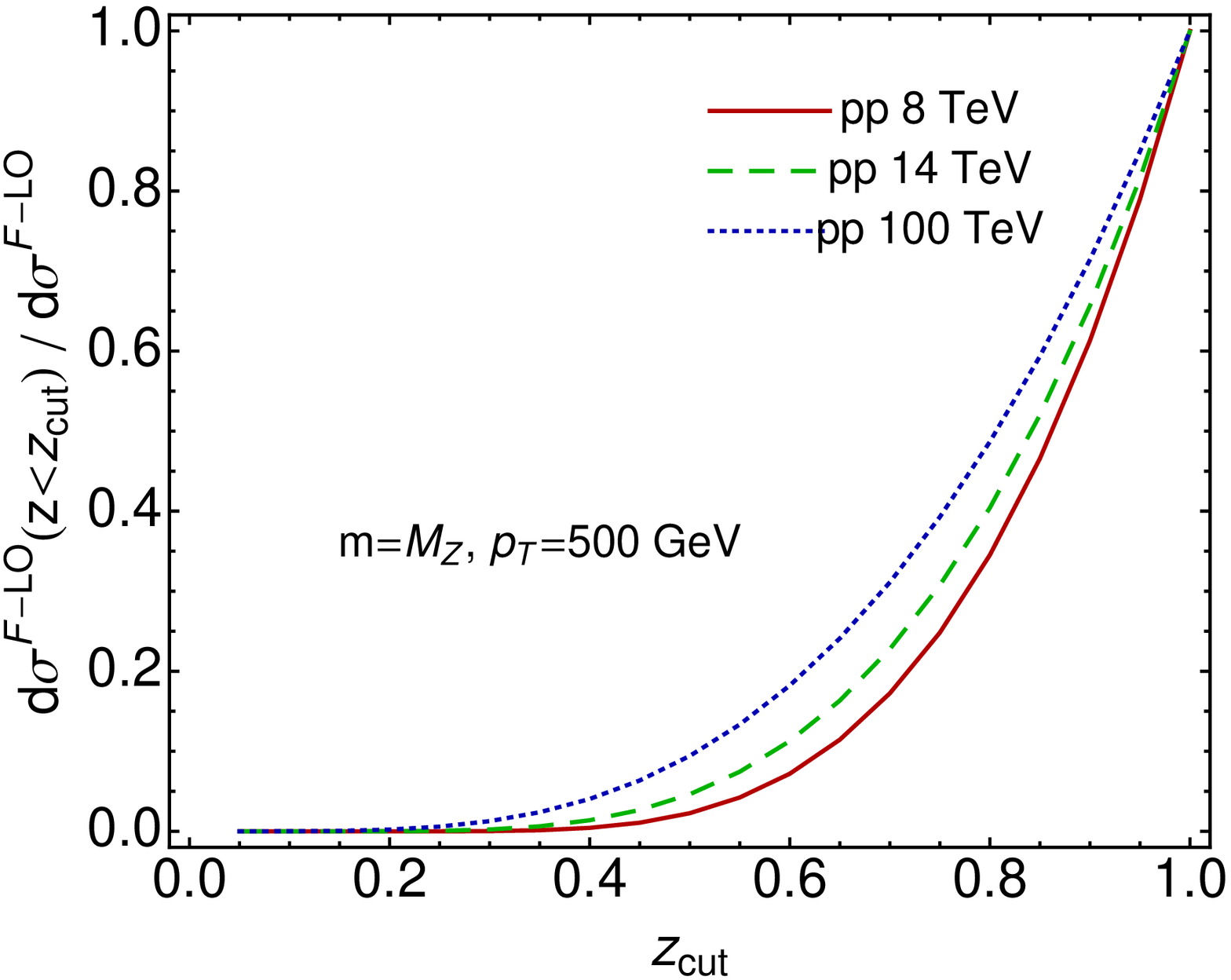}\vspace{0.1in}
  \includegraphics[width=0.45\textwidth]{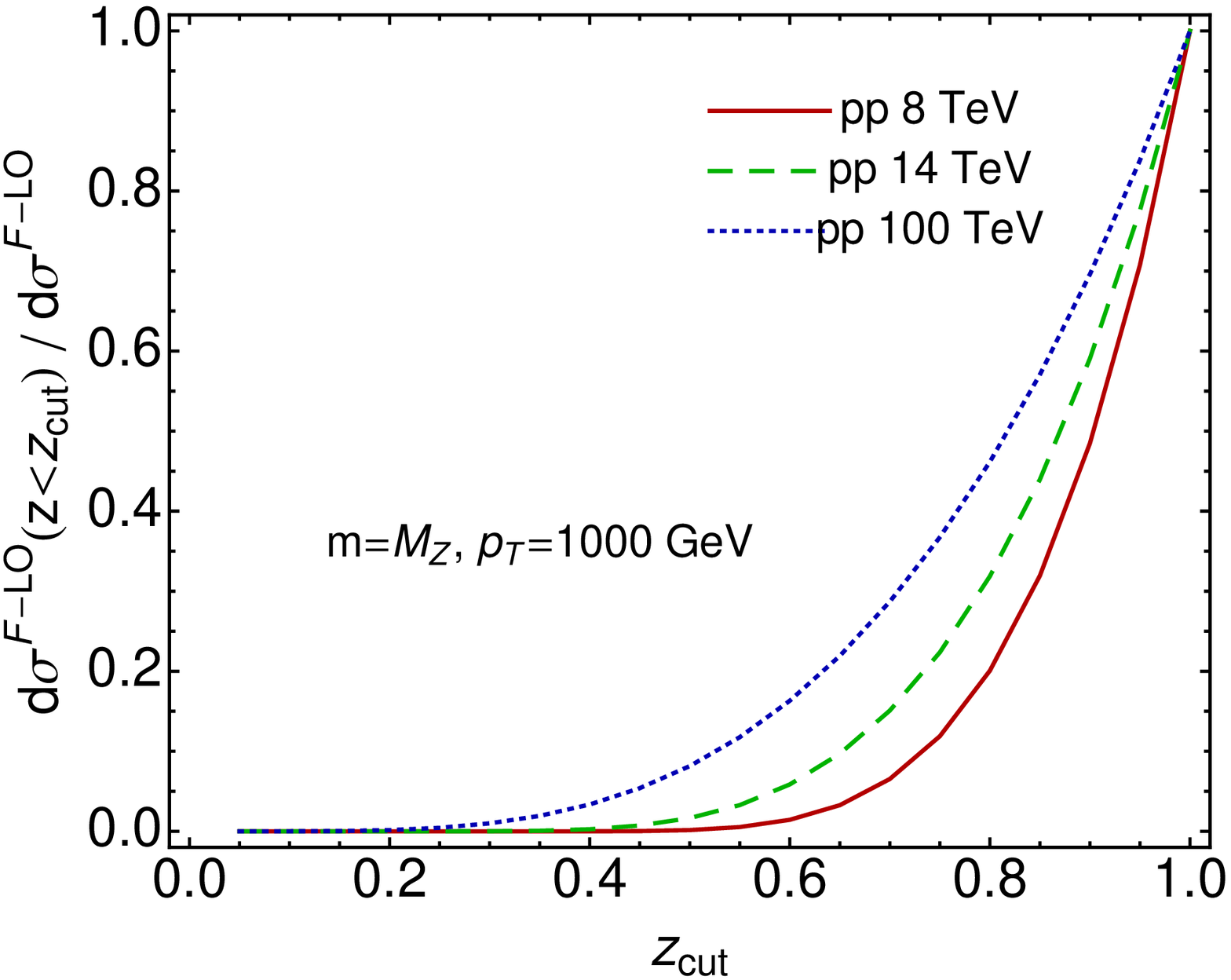}
  \end{center}
  \vspace{-1ex}
  \caption{\label{fig:frag3}
 Fractional contribution of $\sigma^{\rm F-LO}$ for Z boson production 
 as a function of the cutoff, $z_{\rm cut}$, of the upper limit of $z$-integration
 in Eq.~(\ref{DR-F-fac}), at
 $p_T=500$ and $1000\,{\rm GeV}$ and $y=0$. }
\end{figure}

Equation~(\ref{DR-F-fac}) expresses the fragmentation contribution as a 
convolution of two PDFs and one fragmentation function, and the hard scale of 
the partonic scattering is proportional to the combination $x_1\, x_2/z$.  
The factored fragmentation 
contribution should likely be dominated by the kinematic region 
where $x_1$ and $x_2$ are small, while $z$ is relatively large, owing to
the competition of two steeply falling PDFs against one fragmentation 
function when the momentum fractions, $x$ or $z$ increase.  
To better understand the dominant region of partonic scattering, we introduce
a cutoff, $z_{\rm cut} \leq 1$, to limit the $z$-integration in Eq.~(\ref{DR-F-fac})
to $z_{\rm min} \leq z \leq z_{\rm cut}$, instead of $z_{\rm min} \leq z \leq 1$.  
In Fig.~\ref{fig:frag3}, we plot $\sigma^{\rm F-LO}$ for $Z$ boson production 
as a function of the cutoff, $z_{\rm cut}$, normalized by the full contribution with
$z_{\rm cut}=1$, to show the fractional contributions from a limited phase space
in $z$.  As expected, the contributions at large $z$ 
dominates in the fragmentation pieces. For example, the domain $z>0.5$ 
contributes over 90\% of the fragmentation cross
sections for all collision energies shown in Fig.~\ref{fig:frag3}. 
The effective range of $z$ is even more limited to the larger $z$-values
for production at the LHC energies owing to the smaller phase space
for a given $p_T$.

\begin{figure}[h!]
  \begin{center}
  \includegraphics[width=0.45\textwidth]{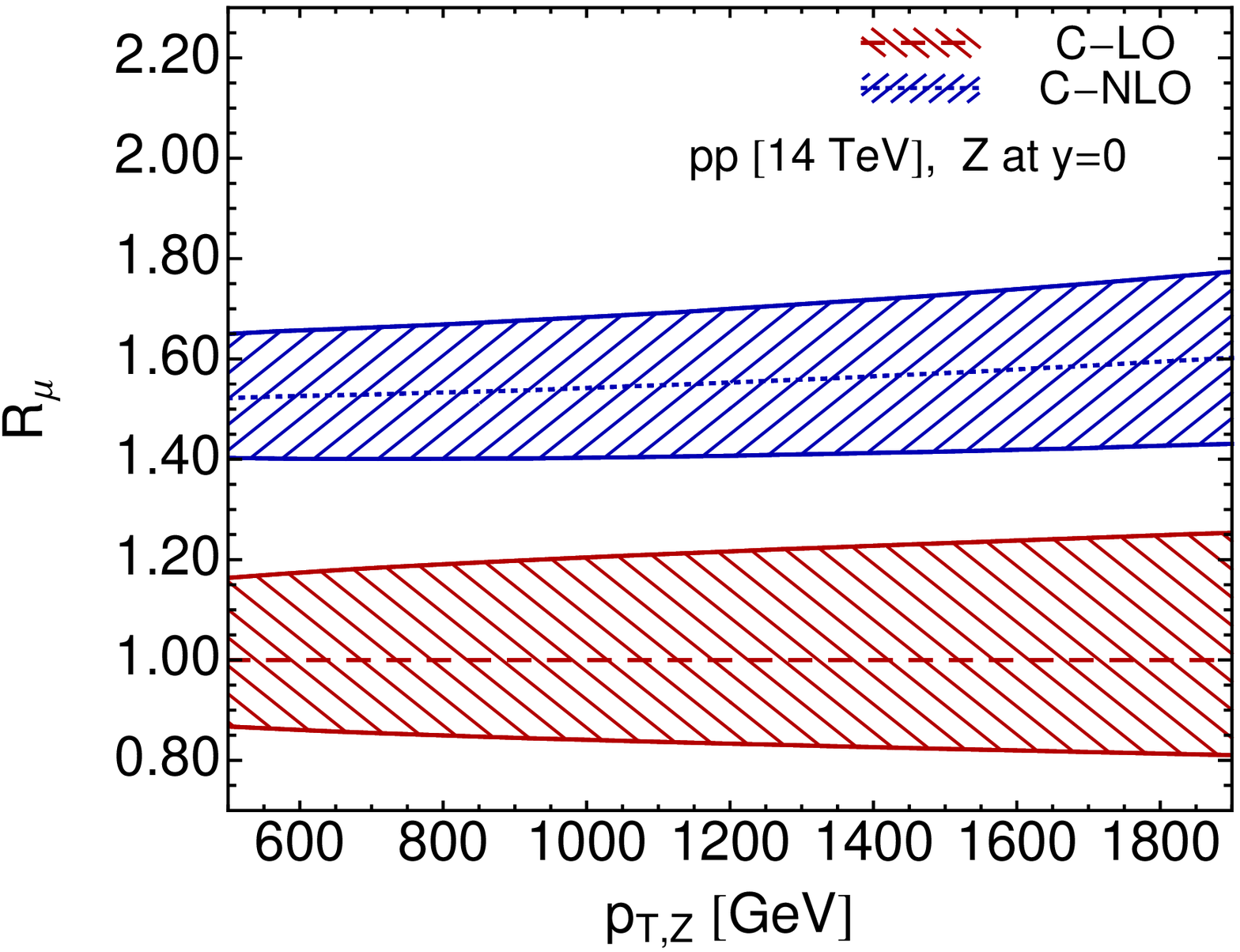}\vspace{0.1in}
  \includegraphics[width=0.45\textwidth]{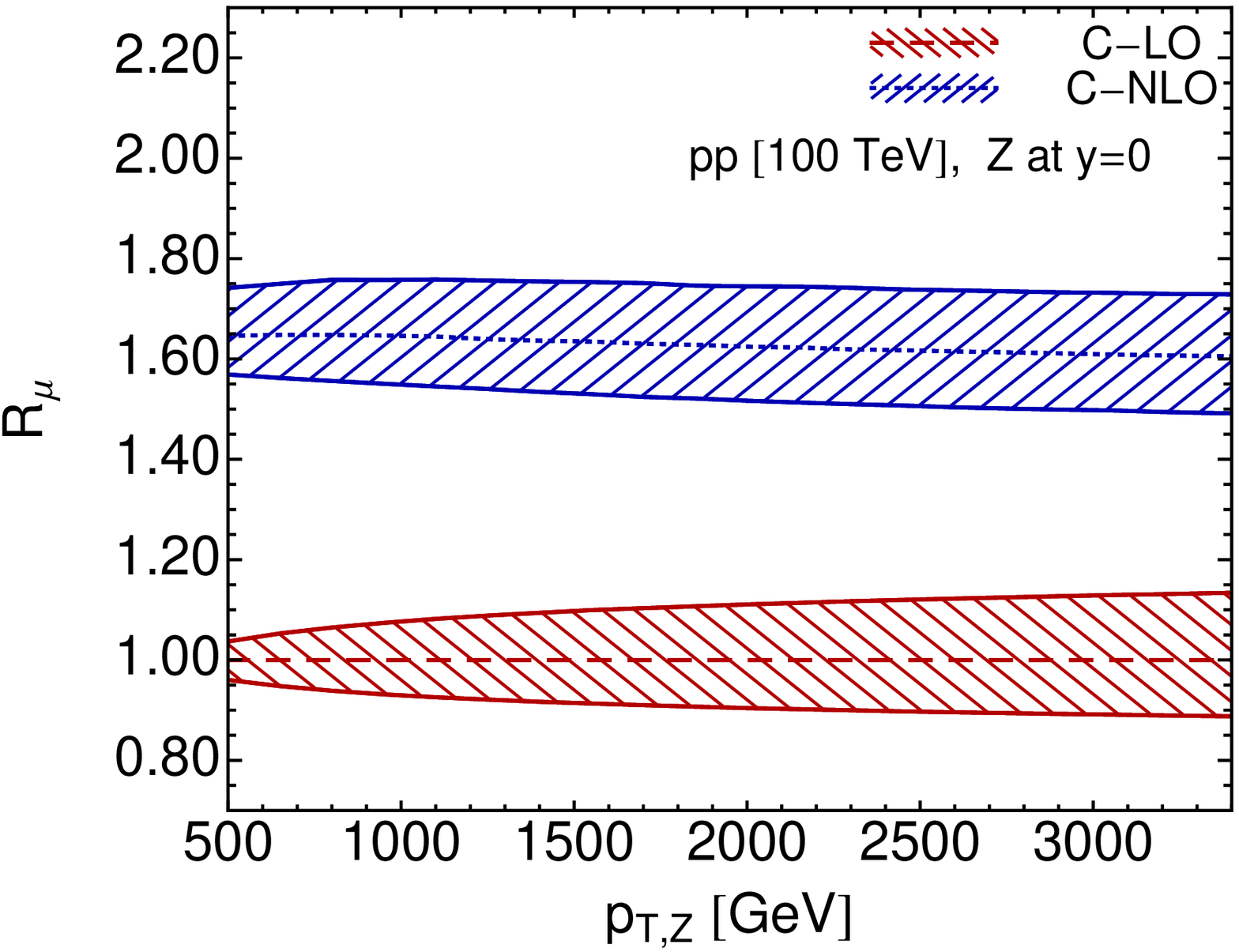}
  \end{center}
  \vspace{-1ex}
  \caption{\label{fig:rat4}
 Conventional LO and NLO predictions on double-differential cross sections of
 $Z$ boson production at 14 TeV and 100 TeV, including
 the central predictions and the scale variations, all normalized to the
 central predictions of the conventional LO results. }
\end{figure}

In Fig.~\ref{fig:rat4}, we plot the LO and NLO contributions to the double-differential cross 
sections of $Z$ boson production at $y=0$, evaluated in the conventional fixed-order perturbative expansion 
and normalized by the conventional LO cross sections.  The uncertainty bands in Fig.~\ref{fig:rat4} 
are calculated by varying the renormalization and factorization scales, 
$\mu_r=\mu_f=\mu$, upward and downward by a factor of two, as an estimation of
the remaining higher-order contributions. 
There are moderate reductions of the scale variation from LO to NLO for the LHC energy, 
while the LO result at the 100 TeV collider shows a scale variation that is too small 
due to accidental cancellations between the renormalization and the factorization scale
dependence.  One may also notice that the 
NLO bands do not overlap the LO bands, suggestive that the scale variations for the conventional 
fixed-order perturbative expansion may underestimate the true theoretical uncertainties from the 
remaining high order corrections.

\subsection{$W$ and $Z$ boson
production at large $p_T$ in the modified expansion scheme}

\begin{figure}[h!]
  \begin{center}
  \includegraphics[width=0.45\textwidth]{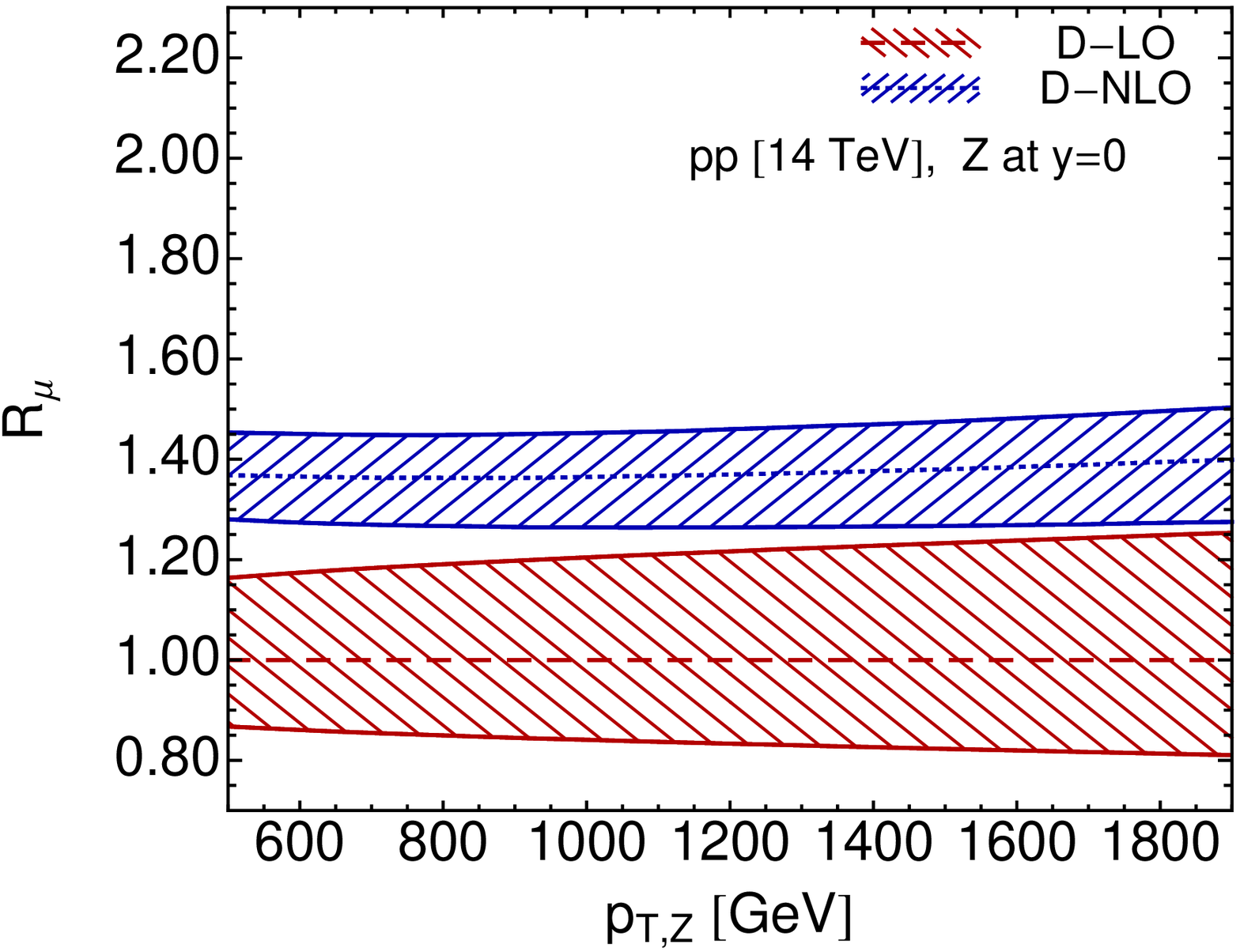}\vspace{0.1in}
  \includegraphics[width=0.45\textwidth]{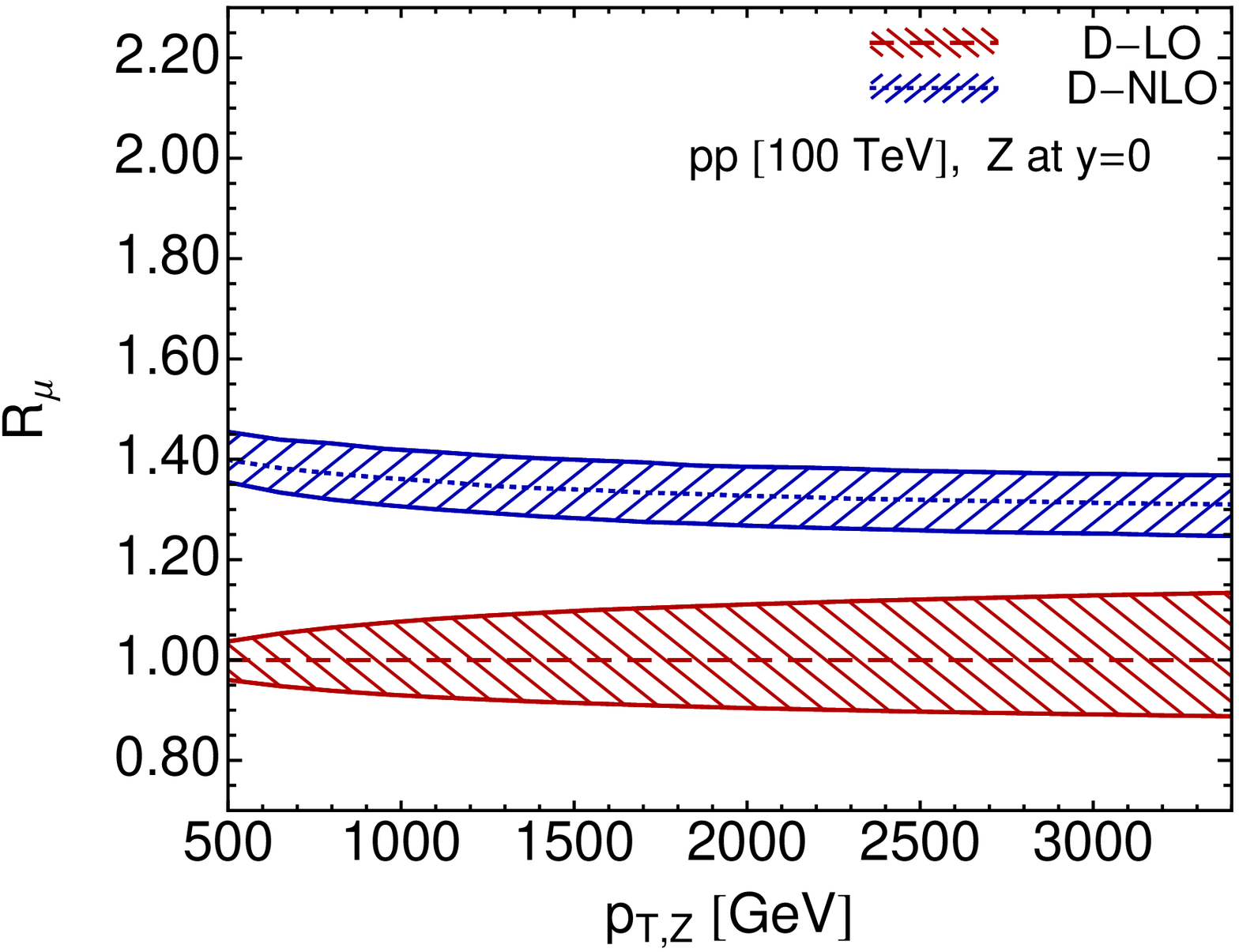}
  \end{center}
  \vspace{-1ex}
  \caption{\label{fig:rat5}
 LO and NLO predictions for the ``direct'' contributions to double-differential cross sections of 
 $Z$ boson production in proton-proton collisions at $\sqrt{s}=14$ TeV and 100 TeV, respectively, 
 with uncertainty bands from the scale variations.  The predictions are normalized by the
 central values of the conventional LO cross sections. }
\end{figure}

\begin{figure}[h!]
  \begin{center}
  \includegraphics[width=0.45\textwidth]{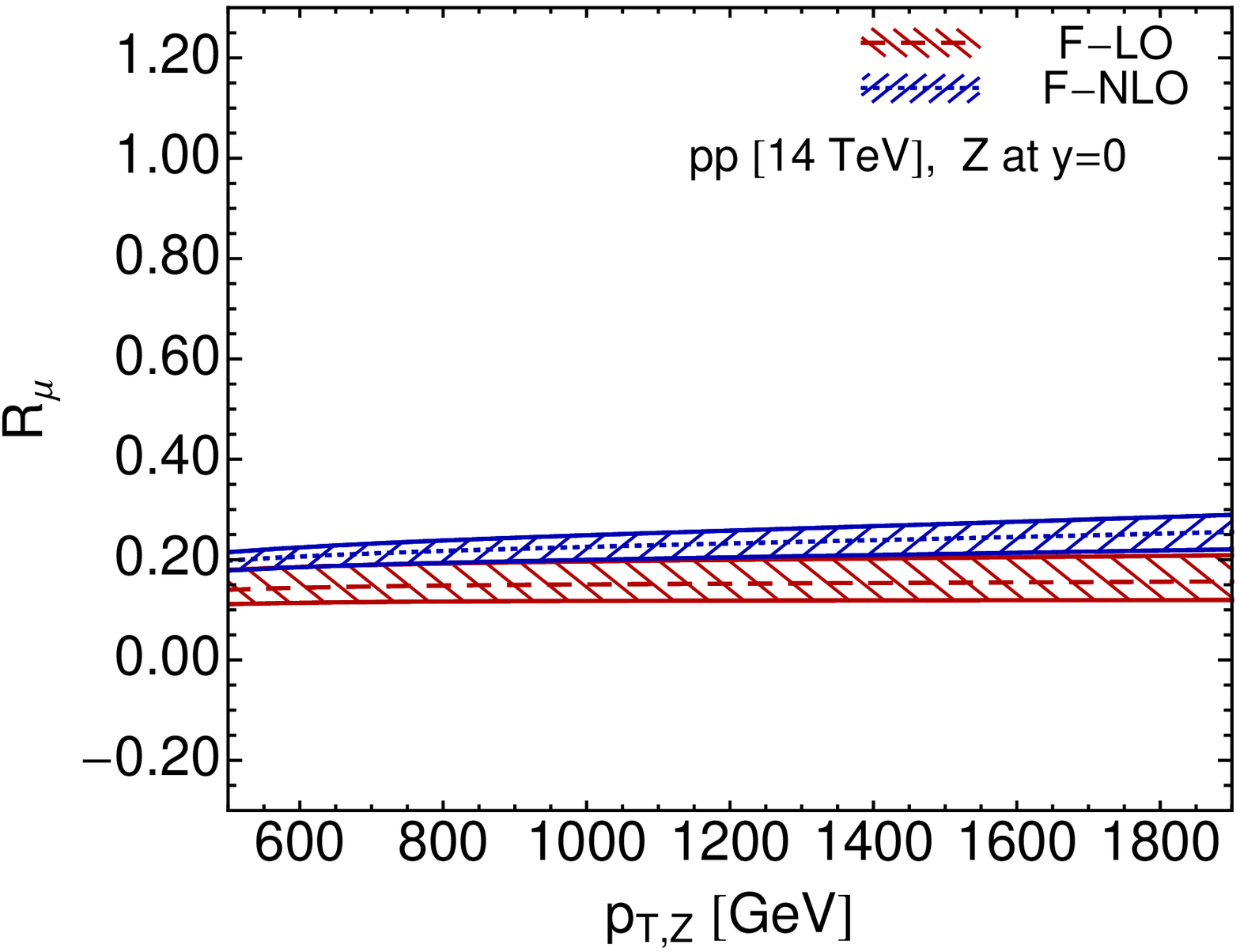}\vspace{0.1in}
  \includegraphics[width=0.45\textwidth]{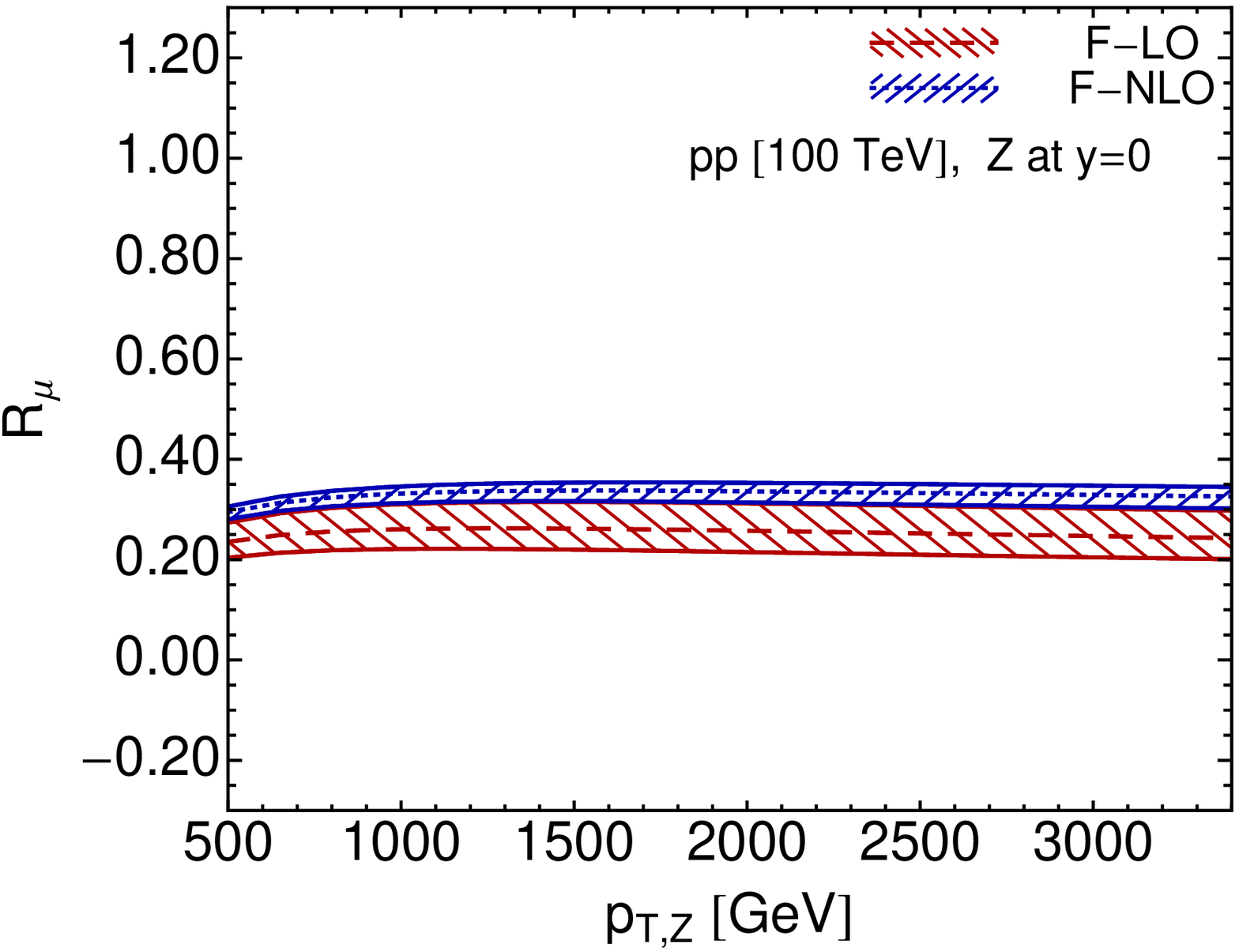}
  \end{center}
  \vspace{-1ex}
  \caption{\label{fig:rat6}
 LO and NLO predictions for the ``fragmentation'' contributions to double-differential cross sections of
 $Z$ boson production in proton-proton collisions at $\sqrt{s}=14$ TeV and 100 TeV, respectively, 
 with uncertainty bands from the scale variations.  The predictions are normalized by the
 central values of the conventional LO cross sections. }
 \end{figure}

In our modified factorization scheme, we split the conventional 
fixed-order perturbative series into the ``direct'' and ``fragmentation'' contributions, 
as shown in Eq.~(\ref{eq:modified-fac}), and we  
calculate them to NLO in this paper.
In Fig.~\ref{fig:rat5}, we plot both the LO and NLO ``direct'' contributions 
to the double-differential cross sections of $Z$ boson production.  The results 
are presented as a ratio ${\rm R}_\mu$, normalized by the conventional LO cross sections.
Note that as mentioned earlier the ``direct'' contribution at LO equals
the conventional LO cross section.    
Similarly, in Fig.~\ref{fig:rat6}, we plot the corresponding 
LO and NLO ``fragmentation'' contributions, 
normalized by the same conventional LO cross sections.
In our modified factorization scheme,  there is a fragmentation scale $\mu_D$, in 
addition to the renormalization scale $\mu_r$ and factorization scale $\mu_f$, which 
are the same as those in the conventional fixed-order factorization scheme. 
Different choices of $\mu_D$ effectively move some finite perturbative contributions
between the ``direct'' and the ``fragmentation'' contributions in Eq.~(\ref{eq:modified-fac}). 
We set the default choice for all three scales equal to $m_T$.
The uncertainty bands in Figs.~\ref{fig:rat5} and \ref{fig:rat6} 
are calculated by varying the renormalization and factorization scales, 
$\mu_r=\mu_f=\mu$, upward and downward by a factor of two, as an estimation of
the remaining higher-order contributions.  The effects from varying 
the fragmentation scale, $\mu_D$, are discussed later. 

Comparing the relative size of NLO corrections in our modified factorization scheme
with those of the conventional fixed-order expansion, we see in 
Figs.~\ref{fig:rat4}, \ref{fig:rat5}, and \ref{fig:rat6}  
that the NLO K-factors in the conventional expansion are significantly 
larger than those evaluated in the modified scheme.
The improved convergence of our modified expansion scheme indicates
a better control of unknown higher-order corrections.   
In Fig.~\ref{fig:rat6}, we observe significant
reduction of the scale variation for the fragmentation contributions at NLO 
as a consequence of the inclusion of the NLO partonic hard parts.

\begin{figure}[h!]
  \begin{center}
  \includegraphics[width=0.45\textwidth]{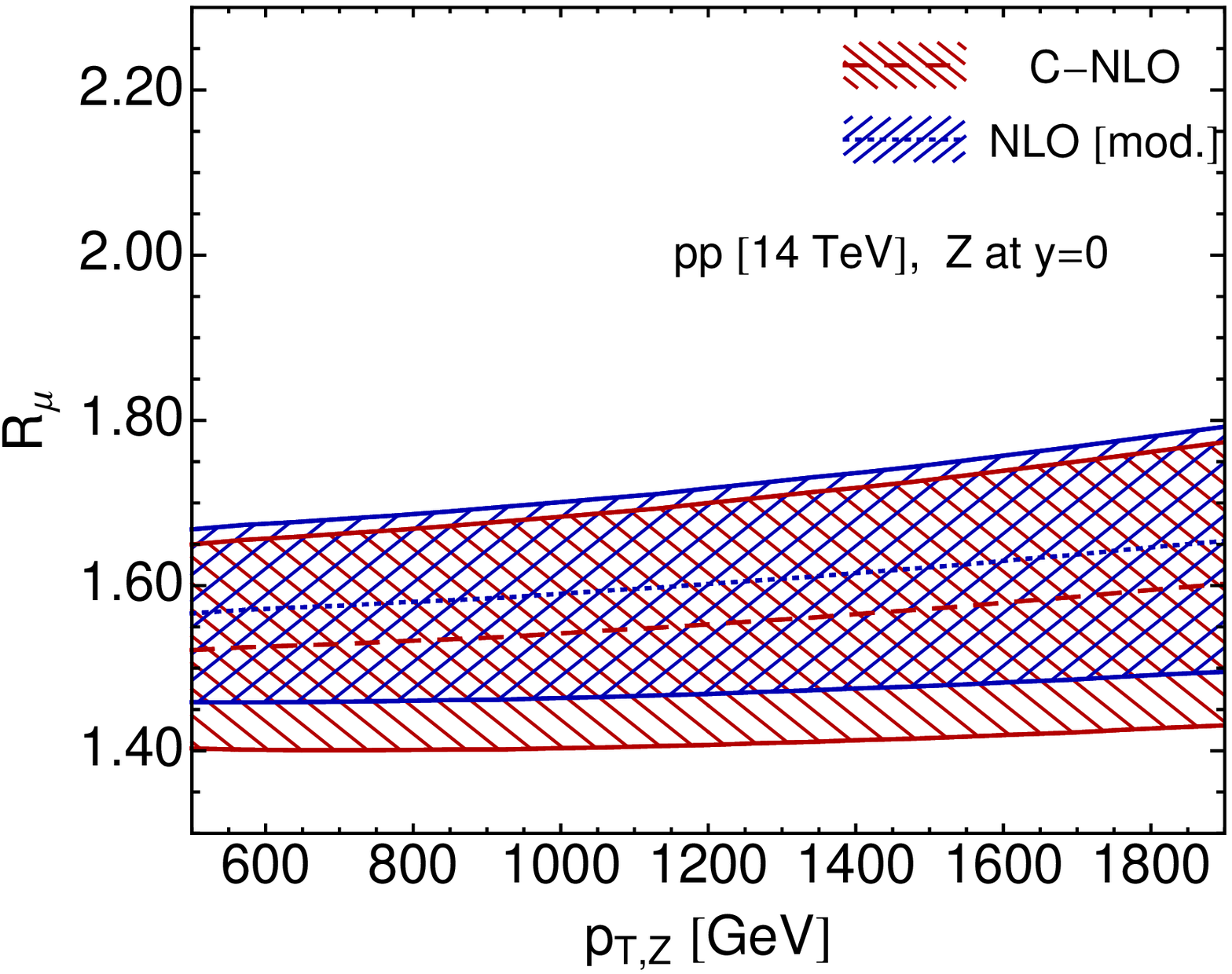}\vspace{0.1in}
  \includegraphics[width=0.45\textwidth]{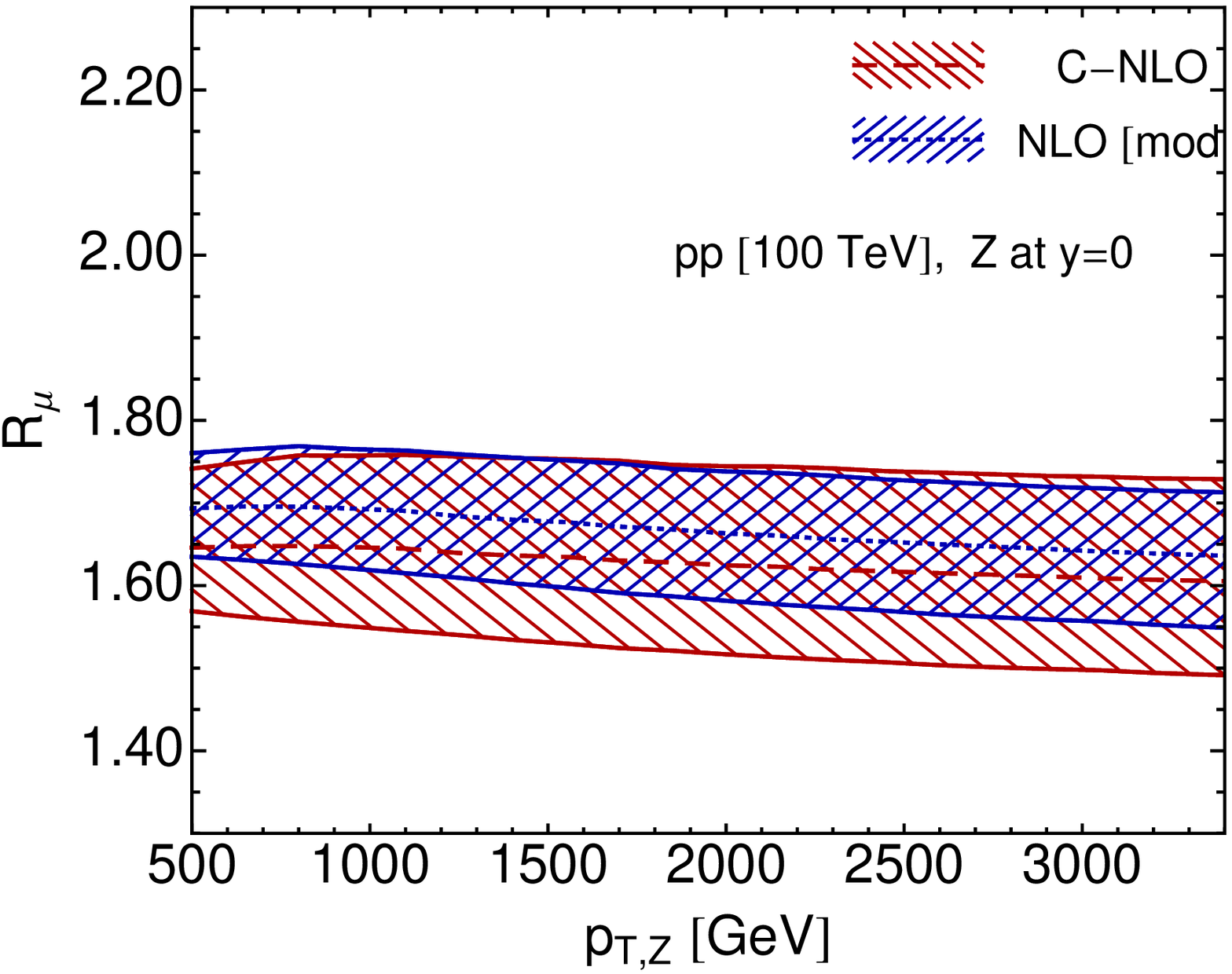}
  \end{center}
  \vspace{-1ex}
  \caption{\label{fig:rat3}
 The full NLO predictions in the modified factorization scheme
 for double-differential cross sections of $Z$ boson production at 
 14 TeV and 100 TeV, including
 the central predictions and the scale variations, together
 with the conventional NLO results, all normalized to the
 central predictions of the conventional LO results. }
\end{figure}

As expressed in Eq.~(\ref{eq:modified-fac}),
the full NLO predictions in our modified factorization scheme 
are equal to the sum of the NLO ``direct'' and ``fragmentation'' 
contributions.  They differ from the conventional NLO results 
by additional higher-order contributions, or the type-B corrections, 
defined in Eq.~(\ref{eq:diff-nlo}). 
In Fig.~\ref{fig:rat3}, we plot the full NLO predictions in our 
modified factorization scheme for the double-differential cross sections 
of $Z$ boson production, together with the conventional NLO results, where we 
show both the central predictions and the scale variation.  Both 
are normalized by the central values 
of the conventional LO cross sections.  Note that the scale-variation  
(blue) band of the modified factorization scheme lies almost entirely 
within the (red) band of the conventional scheme.   
At NLO accuracy, the predictions from the conventional 
fixed-order perturbative expansion show about 20\% uncertainty from
the scale variation, which increases with $p_T$, 
while as expected, the full NLO predictions from our modified 
factorization scheme show a reduction of the uncertainty, 
because of the inclusion of the type-B corrections.  
The reduction is stronger for the NLO predictions at 
$\sqrt{s} = 100$~TeV, where the fragmentation contributions
are significantly larger, as shown in Fig.~\ref{fig:rat1}. 

\begin{figure}[h!]
  \begin{center}
  \includegraphics[width=0.45\textwidth]{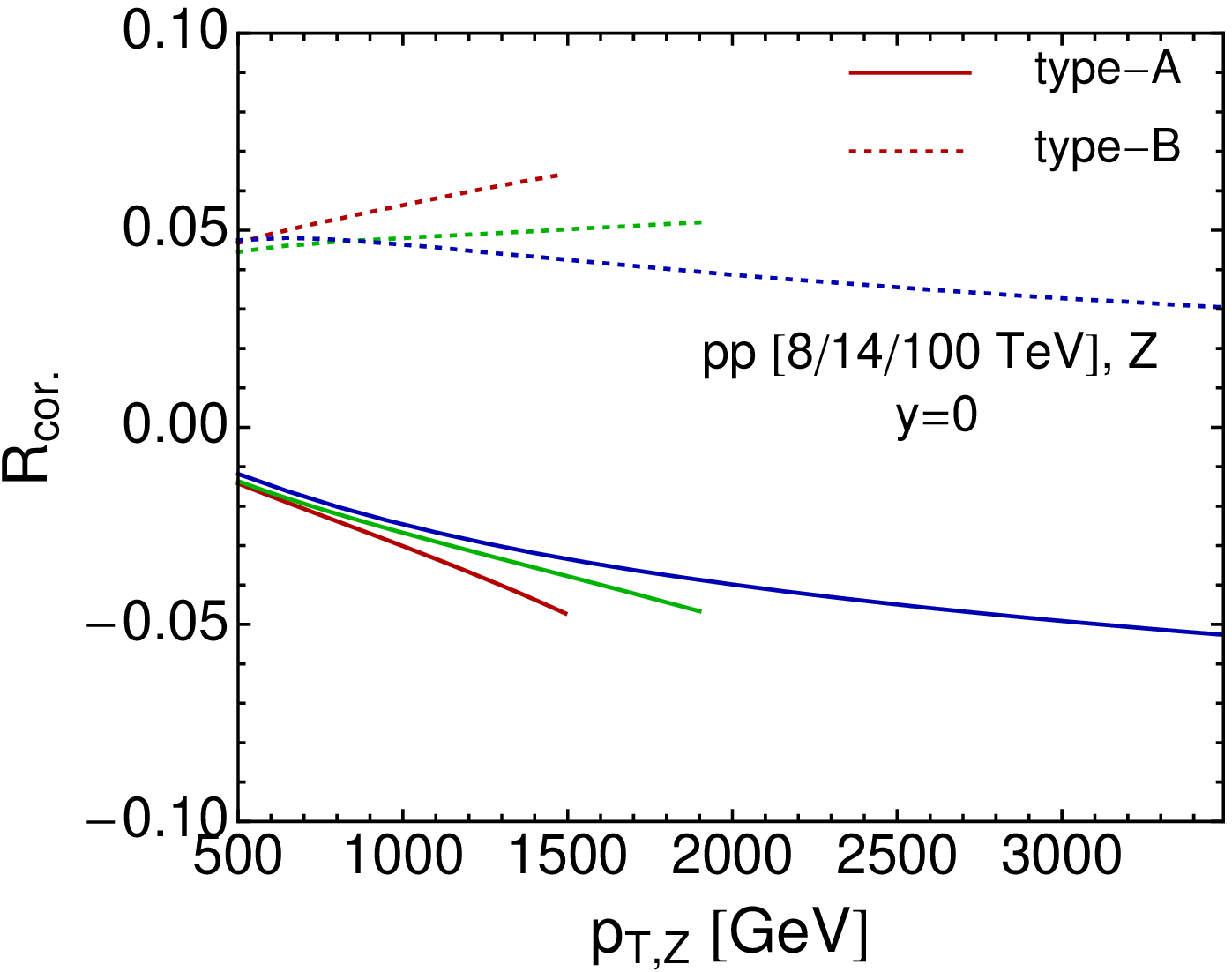}\vspace{0.1in}
  \includegraphics[width=0.45\textwidth]{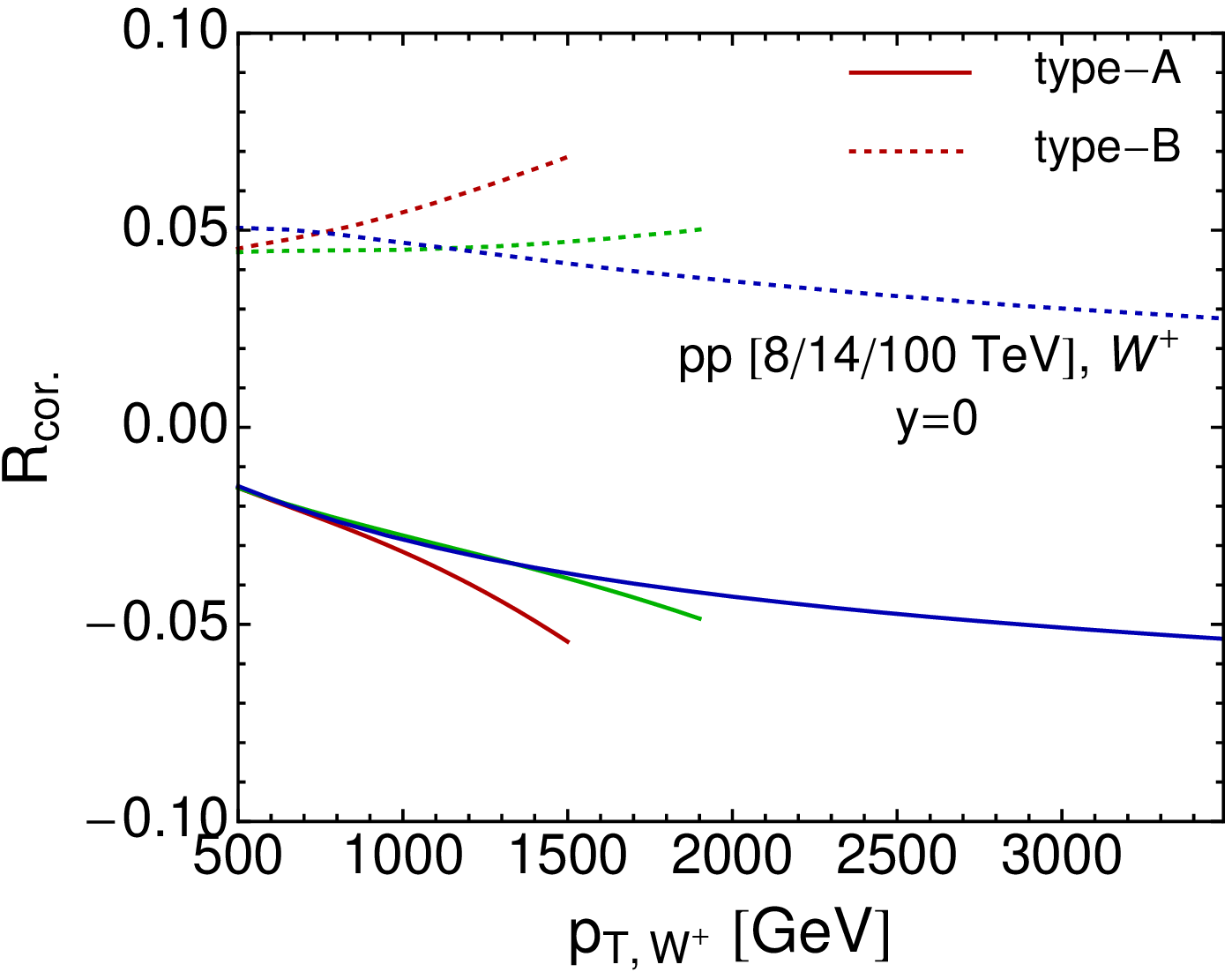}
  \end{center}
  \vspace{-1ex}
  \caption{\label{fig:rat2}
 Differences between the full NLO predictions in the modified factorization scheme
 and the conventional NLO predictions, for double-differential cross sections
 of $Z$ and $W^+$ boson production at 8, 14, and 100 TeV, 
 in each sub-figure with increasing range of $p_T$. }
\end{figure}

The central values of the NLO predictions
from our modified factorization scheme are larger than 
the NLO predictions from the conventional factorization scheme,
as shown in Fig.~\ref{fig:rat3}.  This result 
is an immediate consequence of the type-B corrections in
Eq.~(\ref{eq:diff-nlo}).  To further quantify this difference,
we plot the type-B corrections for both the $Z$ and $W^+$ boson production
in Fig.~\ref{fig:rat2}, along with the type-A corrections, 
as specified in the discussion immediately after Eq.~(\ref{eq:fasym-nlo}).  
Both type-A and type-B corrections in Fig.~\ref{fig:rat2} are shown as ratios 
to the conventional LO cross sections, 
${\rm R}_{\rm cor.} = (\sigma^{\rm D-NLO} + \sigma^{\rm F-NLO} - \sigma^{\rm C-NLO})/
\sigma^{\rm C-LO}$.
Unlike the type-B corrections, which are positive and 
defined to be the total corrections at NLO, the type-A corrections are
negative and are of a few percent of the LO cross sections.
As shown in Eq.~(\ref{eq:diff-nlo}), the type-A corrections 
are proportional to the difference between the fully-resummed 
and the LO perturbative fragmentation functions.  
The fully resummed fragmentation functions have a 
softer $z$ dependence when compared to the LO piece of the functions,
and naturally, the difference is negative for the large $z$ region, while 
it becomes positive and larger as $z$ decreases, as shown in Fig.~\ref{fig:frag2}.  
The negative type-A corrections, as shown in Fig.~\ref{fig:rat2},
arise from the fact that the fragmentation contributions are dominated 
by the large $z$ region, as discussed above.
Numerically, the type-A corrections are within $-6$\% for all cases shown in Fig.~\ref{fig:rat2},
similar to the results for virtual photon production~\cite{Berger:2001wr}. 
At a given $p_T$, the absolute size of the corrections is larger for production  
at a lower $\sqrt{s}$ owing to the smaller phase space for radiation and 
larger effective $z$ values.
The net type-B corrections are positive, despite the negative type-A corrections, 
because the fragmentation corrections from NLO hard matrix elements are positive, 
bringing the total corrections up by 6-10\%.
This outcome is understandable since the NLO corrections to parton or jet production 
at hadron colliders, which share the similar hard matrix elements, 
are usually positive and large. 

As shown in Fig.~\ref{fig:rat2}, the final numerical differences between the NLO predictions,
calculated in our modified factorization scheme and the conventional fixed-order
perturbative expansion are about 5\% 
at $p_T=500\,{\rm GeV}$, and increase or decrease slightly for the case of the LHC 
or a 100 TeV proton-proton collider.   The corrections
are similar for $Z$ and $W^+$ boson production.
Based on the relatively weak $p_T$ dependence of the corrections in Fig.~\ref{fig:rat2}, 
we expect that the NLO predictions from our modified factorization scheme 
will not differ appreciably for the overall shape of the large $p_T$ spectrum of the $W$ or $Z$ 
at the LHC derived from the conventional NLO calculations. 
At the 100 TeV proton-proton collider, our NLO predictions for the large $p_T$ spectrum
of both the $W$ and $Z$ boson production could be slightly softer than 
the NLO spectrum evaluated from conventional fixed-order perturbative 
calculations without resummation of the fragmentation logarithms.  
It is the resummation of the large fragmentation logarithms that makes our 
modified factorization formalism more stable and converge faster perturbatively.

To examine the fragmentation scale dependence of the NLO predictions
of our modified factorization scheme, we increased and decreased the scale
by a factor of two from its default value, $\mu_D=m_T$, while fixing the
renormalization and factorization scales. 
Our NLO results in the modified factorization scheme show a very weak
dependence on $\mu_D$, at a level of 1\%. 
By comparing the fragmentation scale dependence between the ``direct'' 
and the ``fragmentation'' contribution to the production cross sections,  
$\sigma^{\rm D}$ and $\sigma^{\rm F}$ in Eq.~(\ref{eq:modified-fac}), we observe 
that the leading dependence on $\mu_D$ in $\sigma^{\rm F}$ cancels 
that in $\sigma^{\rm D}$ by definition.  
The sub-leading dependence on $\mu_D$ from QCD evolution 
cancels within $\sigma^{\rm F}$ between the resummed corrections to the fragmentation
functions and the NLO corrections to hard matrix elements.
Thus the remaining dependence on $\mu_D$ in the sum of 
$\sigma^{\rm D-NLO}$ and $\sigma^{\rm F-NLO}$ is negligible.

In conclusion, within the modified factorization scheme we split 
vector-boson production at large $p_T$ into two separate perturbation series, 
both of which show better convergence compared to the conventional fixed-order
calculations. By including some higher-order corrections (beyond NLO
in conventional fixed order) for the
fragmentation contributions we reduce the scale variations
and obtain a perturbative expansion with better control of the theoretical
uncertainties, especially for vector-boson production at the 100 TeV collider.

\subsection{Discussion of the Numerical Results}

In Table~\ref{tab:bin} we summarize our total cross sections for 
vector-boson production, integrated  over the full rapidity range, with $p_{T,V}$ 
greater than 800 GeV.  We provide predictions based on conventional LO and 
NLO in the second and third columns, and the NLO predictions using the modified factorization
scheme in the last column, for the LHC at 13 TeV and for a future proton-proton collider at 100 TeV. 
We choose a $p_T$ threshold of 800 GeV for illustration.
As shown in Table~\ref{tab:bin}, the total cross section for $Z$ boson production 
at $\sqrt{s}= 13$ TeV can reach 120 fb with a scale uncertainty about $\pm 8$\%.
The scale uncertainty is about $\pm 5$\% for $Z$ boson production at the 100 TeV collider.
Uncertainties due to variations of the fragmentation
scale $\mu_D$ are not
included since they are much smaller.
These residual scale variations should be further reduced either by 
soft gluon resummation~\cite{Kidonakis:1999ur,Kidonakis:2003xm,Gonsalves:2005ng,Becher:2011fc,Kidonakis:2012sy,Becher:2012xr,Becher:2013vva,Kidonakis:2014zva}
or ongoing NNLO calculations.  Other uncertainties include PDF uncertainties of 
about $\pm 4$\% for $Z$ or $W$ boson production at 13 TeV, and about $\pm 1$\% for 
the 100 TeV collider, obtained using the META PDF
prescription proposed in~\cite{Gao:2013bia}. The PDF uncertainties are much
larger at 13 TeV because the large-$x$ region of the PDFs is being probed, a situation that 
could be improved once more LHC data are included in the global analysis of PDFs.
For vector-boson production at large $p_T$ at the LHC, the electroweak
corrections due to Sudakov logarithms are also 
significant~\cite{Maina:2004rb,Kuhn:2004em,Kuhn:2005az,Hollik:2007sq,Kuhn:2007qc,Kuhn:2007cv,Becher:2013zua,Kallweit:2014xda}
and must be included when comparison is made with data.

\begin{table}[h!]
\centering
\begin{tabular}{l|l|l|l|l} \hline
 $\sigma_{p_T>800\,{\rm GeV}}$ & VB & C-LO & C-NLO & NLO[modified]  \\  [2ex] 
\hline \hline 
\multirow{3}{*}{13 TeV [fb]} & $Z$ & 74.1 & $117.4^{+12.0}_{-11.5}$ & $120.5^{+9.2}_{-10.4}$  \\  [1ex] 
  & $W^+$ &  126.2 & $199.4^{+20.1}_{-19.3}$ & $204.4^{+15.4}_{-17.3}$  \\ [1ex]
  & $W^-$ &  55.8 & $90.2^{+9.6}_{-9.2}$ & $92.7^{+7.4}_{-8.2}$  \\ [1ex]
\hline
\multirow{3}{*}{100 TeV [pb]} & $Z$ & 11.48 & $19.68^{+1.53}_{-1.30}$ & $20.16^{+1.00}_{-0.98}$  \\  [1ex] 
   & $W^+$ & 15.08 & $26.23^{+2.14}_{-1.79}$ & $26.86^{+1.41}_{-1.35}$  \\ [1ex]
  & $W^-$ &  10.50 & $18.18^{+1.47}_{-1.23}$ & $18.61^{+0.96}_{-0.92}$  \\ [1ex]
\hline
\end{tabular}
\caption{Cross sections integrated over $y$ and $p_T>800\,{\rm GeV}$ with scale variations,
for $W$ and $Z$ boson production at 13 TeV (in fb) and at 100 TeV (in pb).
The scale uncertainties are calculated by
varying $\mu_r=\mu_f=\mu$ by a factor of two in both directions around $m_T$.
The uncertainties due to variations of $\mu_D$ are not included since they
are much smaller.}
\label{tab:bin}
\end{table}

Measurements exist at $\sqrt{s}=7$ TeV at the LHC 
of the $Z$ boson $p_T$ spectrum in the leptonic channel~\cite{Aad:2014xaa} 
and of the $Z$ or $W$ boson cross sections above a certain $p_T$ threshold 
in the hadronic channel~\cite{Aad:2014haa}.
The highest effective $p_T$ of the vector boson is limited to about 300 GeV.  At 13 TeV, 
according to Table~\ref{tab:bin}, $\sim$2400 $Z$ bosons are expected with $p_T$ 
above 800 GeV in the di-muon and di-electron channel, assuming an integrated 
luminosity of 300 ${\rm fb}^{-1}$.  The statistical uncertainties are estimated to be
smaller than the scale variations in Table~\ref{tab:bin}.  Thus precise 
tests of various QCD predictions require good control of the experimental systematical 
uncertainties at 13 TeV and 100 TeV.
The situation is similar for $W$ boson production where we gain a larger
branching ratio in the leptonic channel but suffer from lower
efficiency in the event reconstruction.

\section{Summary}
\label{sec:con}

We introduced a modified factorization scheme for evaluating the 
transverse momentum spectrum of the SM $W$ and $Z$ boson production 
at hadron colliders. 
In this new scheme we re-organize the conventional 
fixed-order QCD perturbative expansion into two separate perturbative expansions, 
corresponding to ``direct'' and ``fragmentation'' contributions. 
The fragmentation piece has a typical two-stage pattern, production of an on-shell parton 
convoluted with the perturbative
fragmentation functions of the $W$ and $Z$ boson, as illustrated in Fig.~\ref{fig:fflogs}. 
When $p_T^2\gg Q^2$, the large perturbative fragmentation logarithms of $\ln(p_T^2/Q^2)$ 
are resummed into the fragmentation functions by solving their evolution equations.
Consequently, the short-distance partonic hard parts for both the ``direct'' and 
the ``fragmentation'' contributions are free of these large fragmentation logarithms,
and the overall convergence of the re-organized perturbative expansions is 
improved, in particular, for the region where $p_T^2\gg Q^2$.

In our explicit NLO calculations, the fragmentation logarithms make up a large
portion of the conventional NLO K-factors at high $p_T$ at the LHC and especially at a future
100 TeV collider.  Our improved NLO predictions retain all ingredients in the
conventional NLO calculations (up to ${\mathcal O}(\alpha_{em}\as^2)$), but they 
include partial higher-order corrections.  In comparison with the conventional
fixed-order perturbative expansion, the modified NLO predictions show 
a moderate reduction of the scale variation at large $p_T$, and 
improved convergence.  
The improved NLO predictions are about $5\%$ higher in the normalization  
of the large-$p_T$ spectrum, but they provide only modest changes in the shape of the spectrum.  
In subsequent research, it would be desirable to extend the analysis to NNLO.  
Our modified factorization scheme could be applied to the production of other heavy 
particles, such as the Higgs boson and top quark, at large $p_T$ 
when the $p_T$ of the produced state is much larger than its mass.\\

\begin{acknowledgments}

The research of E.~L.~Berger and J.~Gao in the High Energy Physics Division at
Argonne is supported by the U.~S.\ Department of Energy, High Energy Physics,
Office of Science, under
Contract No.\ DE-AC02-06CH11357.
The research of Z.~B.~Kang is partially supported by the U.S. Department of Energy,
Office of Science, under contract No.~DE-AC52-06NA25396.
The research of J.~W.~Qiu is supported in part by the
U. S. Department of Energy under Contract  No. DE-AC02-98CH10886,
and the National Science Foundation under Grants No. PHY-0969739
and No. PHY-1316617.
H. Zhang is supported by the U.S. DOE under Contracts No. DE-FG02-91ER40618 and DE- SC0011702.

\end{acknowledgments}
%


\begin{thebibliography}{999}

\bibitem{Aad:2012tfa} 
  G.~Aad {\it et al.}  [ATLAS Collaboration],
  Phys.\ Lett.\ B {\bf 716}, 1 (2012)
  [arXiv:1207.7214 [hep-ex]].

\bibitem{Chatrchyan:2012ufa} 
  S.~Chatrchyan {\it et al.}  [CMS Collaboration],
  Phys.\ Lett.\ B {\bf 716}, 30 (2012)
  [arXiv:1207.7235 [hep-ex]].

\bibitem{Berger:1988tu} 
  E.~L.~Berger, F.~Halzen, C.~S.~Kim and S.~Willenbrock,
  Phys.\ Rev.\ D {\bf 40}, 83 (1989)
  [Erratum-ibid.\ D {\bf 40}, 3789 (1989)].

  
\bibitem{Kim:1990kt} 
  C.~S.~Kim, A.~D.~Martin and W.~J.~Stirling,
  Phys.\ Rev.\ D {\bf 42}, 952 (1990).

\bibitem{Malik:2013kba} 
  S.~A.~Malik and G.~Watt,
  JHEP {\bf 1402}, 025 (2014)
  [arXiv:1304.2424 [hep-ph]].

\bibitem{Aad:2014xaa} 
  G.~Aad {\it et al.}  [ATLAS Collaboration],
  JHEP {\bf 1409}, 145 (2014)
  [arXiv:1406.3660 [hep-ex]].

\bibitem{Aad:2014haa} 
  G.~Aad {\it et al.}  [ATLAS Collaboration],
  New J.\ Phys.\  {\bf 16}, no. 11, 113013 (2014)
  [arXiv:1407.0800 [hep-ex]].

\bibitem{Gonsalves:1989ar} 
  R.~J.~Gonsalves, J.~Pawlowski and C.~F.~Wai,
  Phys.\ Rev.\ D {\bf 40}, 2245 (1989).

\bibitem{Baer:1991qf} 
  H.~Baer and M.~H.~Reno,
  Phys.\ Rev.\ D {\bf 44}, 3375 (1991).

\bibitem{Arnold:1990yk} 
  P.~B.~Arnold and R.~P.~Kauffman,
  Nucl.\ Phys.\ B {\bf 349}, 381 (1991).

\bibitem{Maina:2004rb} 
  E.~Maina, S.~Moretti and D.~A.~Ross,
  Phys.\ Lett.\ B {\bf 593}, 143 (2004)
  [Erratum-ibid.\ B {\bf 614}, 216 (2005)]
  [hep-ph/0403050].

\bibitem{Kuhn:2004em} 
  J.~H.~Kuhn, A.~Kulesza, S.~Pozzorini and M.~Schulze,
  Phys.\ Lett.\ B {\bf 609}, 277 (2005)
  [hep-ph/0408308].

\bibitem{Kuhn:2005az} 
  J.~H.~Kuhn, A.~Kulesza, S.~Pozzorini and M.~Schulze,
  Nucl.\ Phys.\ B {\bf 727}, 368 (2005)
  [hep-ph/0507178].

\bibitem{Hollik:2007sq} 
  W.~Hollik, T.~Kasprzik and B.~A.~Kniehl,
  Nucl.\ Phys.\ B {\bf 790}, 138 (2008)
  [arXiv:0707.2553 [hep-ph]].

\bibitem{Kuhn:2007qc} 
  J.~H.~Kuhn, A.~Kulesza, S.~Pozzorini and M.~Schulze,
  Phys.\ Lett.\ B {\bf 651}, 160 (2007)
  [hep-ph/0703283 [HEP-PH]].

\bibitem{Kuhn:2007cv} 
  J.~H.~Kuhn, A.~Kulesza, S.~Pozzorini and M.~Schulze,
  Nucl.\ Phys.\ B {\bf 797}, 27 (2008)
  [arXiv:0708.0476 [hep-ph]].

\bibitem{Becher:2013zua} 
  T.~Becher and X.~Garcia i Tormo,
  Phys.\ Rev.\ D {\bf 88}, no. 1, 013009 (2013)
  [arXiv:1305.4202 [hep-ph]].

\bibitem{Kallweit:2014xda} 
  S.~Kallweit, J.~M.~Lindert, P.~Maierhöfer, S.~Pozzorini and M.~Schönherr,
  arXiv:1412.5157 [hep-ph].

\bibitem{Collins:1984kg} 
  J.~C.~Collins, D.~E.~Soper and G.~F.~Sterman,
  Nucl.\ Phys.\ B {\bf 250}, 199 (1985).

\bibitem{Davies:1984hs} 
  C.~T.~H.~Davies and W.~J.~Stirling,
  Nucl.\ Phys.\ B {\bf 244}, 337 (1984).

\bibitem{Balazs:1997xd} 
  C.~Balazs and C.~P.~Yuan,
  Phys.\ Rev.\ D {\bf 56}, 5558 (1997)
  [hep-ph/9704258].

\bibitem{Ellis:1997ii} 
  R.~K.~Ellis and S.~Veseli,
  Nucl.\ Phys.\ B {\bf 511}, 649 (1998)
  [hep-ph/9706526].

\bibitem{Qiu:2000ga} 
  J.~W.~Qiu and X.~f.~Zhang,
  Phys.\ Rev.\ Lett.\  {\bf 86}, 2724 (2001)
  [hep-ph/0012058].

\bibitem{Qiu:2000hf} 
  J.~W.~Qiu and X.~f.~Zhang,
  Phys.\ Rev.\ D {\bf 63}, 114011 (2001)
  [hep-ph/0012348].

\bibitem{Berger:2002ut} 
  E.~L.~Berger and J.~W.~Qiu,
  Phys.\ Rev.\ D {\bf 67}, 034026 (2003)
  [hep-ph/0210135].
  
\bibitem{Landry:2002ix} 
  F.~Landry, R.~Brock, P.~M.~Nadolsky and C.~P.~Yuan,
  Phys.\ Rev.\ D {\bf 67}, 073016 (2003)
  [hep-ph/0212159].

\bibitem{Mantry:2010mk} 
  S.~Mantry and F.~Petriello,
  Phys.\ Rev.\ D {\bf 83}, 053007 (2011)
  [arXiv:1007.3773 [hep-ph]].

\bibitem{Becher:2010tm} 
  T.~Becher and M.~Neubert,
  Eur.\ Phys.\ J.\ C {\bf 71}, 1665 (2011)
  [arXiv:1007.4005 [hep-ph]].

\bibitem{Bozzi:2010xn} 
  G.~Bozzi, S.~Catani, G.~Ferrera, D.~de Florian and M.~Grazzini,
  Phys.\ Lett.\ B {\bf 696}, 207 (2011)
  [arXiv:1007.2351 [hep-ph]].

\bibitem{Mantry:2010bi} 
  S.~Mantry and F.~Petriello,
  Phys.\ Rev.\ D {\bf 84}, 014030 (2011)
  [arXiv:1011.0757 [hep-ph]].

\bibitem{Becher:2011xn} 
  T.~Becher, M.~Neubert and D.~Wilhelm,
  JHEP {\bf 1202}, 124 (2012)
  [arXiv:1109.6027 [hep-ph]].

\bibitem{Catani:2013tia} 
  S.~Catani, L.~Cieri, D.~de Florian, G.~Ferrera and M.~Grazzini,
  Nucl.\ Phys.\ B {\bf 881}, 414 (2014)
  [arXiv:1311.1654 [hep-ph]].

\bibitem{Wang:2013qua} 
  Y.~Wang, C.~S.~Li, Z.~L.~Liu, D.~Y.~Shao and H.~T.~Li,
  Phys.\ Rev.\ D {\bf 88}, 114017 (2013)
  [arXiv:1307.7520].

\bibitem{Kidonakis:1999ur} 
  N.~Kidonakis and V.~Del Duca,
  Phys.\ Lett.\ B {\bf 480}, 87 (2000)
  [hep-ph/9911460].

\bibitem{Kidonakis:2003xm} 
  N.~Kidonakis and A.~Sabio Vera,
  JHEP {\bf 0402}, 027 (2004)
  [hep-ph/0311266].

\bibitem{Gonsalves:2005ng} 
  R.~J.~Gonsalves, N.~Kidonakis and A.~Sabio Vera,
  Phys.\ Rev.\ Lett.\  {\bf 95}, 222001 (2005)
  [hep-ph/0507317].

\bibitem{Becher:2011fc} 
  T.~Becher, C.~Lorentzen and M.~D.~Schwartz,
  Phys.\ Rev.\ Lett.\  {\bf 108}, 012001 (2012)
  [arXiv:1106.4310 [hep-ph]].

\bibitem{Kidonakis:2012sy} 
  N.~Kidonakis and R.~J.~Gonsalves,
  Phys.\ Rev.\ D {\bf 87}, no. 1, 014001 (2013)
  [arXiv:1201.5265 [hep-ph]].

\bibitem{Becher:2012xr} 
  T.~Becher, C.~Lorentzen and M.~D.~Schwartz,
  Phys.\ Rev.\ D {\bf 86}, 054026 (2012)
  [arXiv:1206.6115 [hep-ph]].

\bibitem{Becher:2013vva} 
  T.~Becher, G.~Bell, C.~Lorentzen and S.~Marti,
  JHEP {\bf 1402}, 004 (2014)
  [arXiv:1309.3245 [hep-ph]].

\bibitem{Kidonakis:2014zva} 
  N.~Kidonakis and R.~J.~Gonsalves,
  Phys.\ Rev.\ D {\bf 89}, no. 9, 094022 (2014)
  [arXiv:1404.4302 [hep-ph]].

\bibitem{Boughezal:2013uia} 
  R.~Boughezal, F.~Caola, K.~Melnikov, F.~Petriello and M.~Schulze,
  JHEP {\bf 1306}, 072 (2013)
  [arXiv:1302.6216 [hep-ph]].

\bibitem{Chen:2014gva} 
  X.~Chen, T.~Gehrmann, E.~W.~N.~Glover and M.~Jaquier,
  Phys.\ Lett.\ B {\bf 740}, 147 (2015)
  [arXiv:1408.5325 [hep-ph]].

\bibitem{Boughezal:2015dva} 
  R.~Boughezal, C.~Focke, X.~Liu and F.~Petriello,
  arXiv:1504.02131 [hep-ph].

\bibitem{Rubin:2010xp} 
  M.~Rubin, G.~P.~Salam and S.~Sapeta,
  JHEP {\bf 1009}, 084 (2010)
  [arXiv:1006.2144 [hep-ph]].


\bibitem{Berger:1998ev} 
  E.~L.~Berger, L.~E.~Gordon and M.~Klasen,
  Phys.\ Rev.\ D {\bf 58}, 074012 (1998)
  [hep-ph/9803387].

\bibitem{Berger:2001wr} 
  E.~L.~Berger, J.~W.~Qiu and X.~f.~Zhang,
  Phys.\ Rev.\ D {\bf 65}, 034006 (2002)
  [hep-ph/0107309].

\bibitem{Kang:2008wv} 
  Z.~B.~Kang, J.~W.~Qiu and W.~Vogelsang,
  Phys.\ Rev.\ D {\bf 79}, 054007 (2009)
  [arXiv:0811.3662 [hep-ph]].

\bibitem{Ma:2014svb} 
  Y.~Q.~Ma, J.~W.~Qiu, G.~Sterman and H.~Zhang,
  Phys.\ Rev.\ Lett.\  {\bf 113}, no. 14, 142002 (2014)
  [arXiv:1407.0383 [hep-ph]].

\bibitem{Azimov:1982ef} 
  Y.~I.~Azimov, Y.~L.~Dokshitzer and V.~A.~Khoze,
  Sov.\ J.\ Nucl.\ Phys.\  {\bf 36}, 878 (1982)
  [Yad.\ Fiz.\  {\bf 36}, 1510 (1982)]; 
  Y.~I.~Azimov, Y.~L.~Dokshitzer, V.~A.~Khoze and S.~I.~Troian,
  Yad.\ Fiz.\  {\bf 40}, 777 (1984)
  [Sov.\ J.\ Nucl.\ Phys.\  {\bf 40}, 498 (1984)]; 
  Y.~L.~Dokshitzer, V.~A.~Khoze and S.~I.~Troian,
  J.\ Phys.\ G {\bf 17}, 1481 (1991).

\bibitem{Collins:1989gx} 
  J.~C.~Collins, D.~E.~Soper and G.~F.~Sterman,
  Adv.\ Ser.\ Direct.\ High Energy Phys.\  {\bf 5}, 1 (1988)
  [hep-ph/0409313].


\bibitem{Qiu:2001nr} 
  J.~W.~Qiu and X.~f.~Zhang,
  Phys.\ Rev.\ D {\bf 64}, 074007 (2001)
  [hep-ph/0101004].

\bibitem{Aversa:1988vb}
F.~Aversa, P.~Chiappetta, M.~Greco, and J.~Guillet,
Nucl.\ Phys.\ {\bf B327}, 105 (1989).

\bibitem{Collins:1988wj} 
  J.~C.~Collins and J.~W.~Qiu,
  Phys.\ Rev.\ D {\bf 39}, 1398 (1989).

\bibitem{Braaten:2001sz} 
  E.~Braaten and J.~Lee,
  Phys.\ Rev.\ D {\bf 65}, 034005 (2002)
  [hep-ph/0102130].



\bibitem{Denner:1990ns} 
  A.~Denner and T.~Sack,
  Nucl.\ Phys.\ B {\bf 358}, 46 (1991).


\bibitem{Beringer:1900zz} 
  J.~Beringer {\it et al.}  [Particle Data Group Collaboration],
  Phys.\ Rev.\ D {\bf 86}, 010001 (2012).

\bibitem{Gao:2013bia} 
  J.~Gao and P.~Nadolsky,
  JHEP {\bf 1407}, 035 (2014)
  [arXiv:1401.0013 [hep-ph]].


\bibitem{Gao:2013xoa} 
  J.~Gao, M.~Guzzi, J.~Huston, H.~L.~Lai, Z.~Li, P.~Nadolsky, J.~Pumplin and D.~Stump {\it et al.},
  Phys.\ Rev.\ D {\bf 89}, no. 3, 033009 (2014)
  [arXiv:1302.6246 [hep-ph]].

\bibitem{Martin:2009iq} 
  A.~D.~Martin, W.~J.~Stirling, R.~S.~Thorne and G.~Watt,
  Eur.\ Phys.\ J.\ C {\bf 63}, 189 (2009)
  [arXiv:0901.0002 [hep-ph]].

\bibitem{Ball:2012cx} 
  R.~D.~Ball, V.~Bertone, S.~Carrazza, C.~S.~Deans, L.~Del Debbio, S.~Forte, A.~Guffanti and N.~P.~Hartland {\it et al.},
  Nucl.\ Phys.\ B {\bf 867}, 244 (2013)
  [arXiv:1207.1303 [hep-ph]].

\end{thebibliography}
\end{document}